\documentclass[printer]{aa}
\usepackage{booktabs,caption}
\usepackage[flushleft]{threeparttable}
\usepackage{dcolumn}
\usepackage{longtable}
\usepackage{threeparttablex}
\usepackage{adjustbox}

\usepackage{graphicx}
\usepackage{natbib}

\usepackage{amsmath}
\usepackage{amssymb}

\usepackage{url}
\usepackage{hyperref}
\usepackage{xcolor}
\definecolor{dark-red}{rgb}{0.4,0.15,0.15}
\definecolor{dark-blue}{rgb}{0.15,0.15,0.4}
\definecolor{medium-blue}{rgb}{0,0,0.5}
\hypersetup{
    colorlinks, linkcolor={dark-red},
    citecolor={dark-blue}, urlcolor={medium-blue}
}

\newcommand{\beqa}{\begin{eqnarray}} 
\newcommand{\eeqa}{\end{eqnarray}}

\newcommand{\bsub}{\begin{subequations}}
\newcommand{\esub}{\end{subequations}}
\newcommand{\beal}{\begin{align}}
\newcommand{\ealn}{\end{align}}
\newcommand{\Nif}{$\rm ^{56}Ni$}

\newcommand{\msun}{M$_{\odot}$}

\newcommand{\Msun}{{\ensuremath{\mathrm{M}_{\odot}}}}

\begin{document}
\title{Oxygen and helium in stripped-envelope supernovae}
\author{C.~Fremling\inst{1,2}\and
J.~Sollerman\inst{1}\and
M.~M.~Kasliwal\inst{2}\and
S.~R.~Kulkarni\inst{2}\and
C.~Barbarino\inst{1}\and
M.~Ergon\inst{1}\and
E.~Karamehmetoglu\inst{1}\and
F.~Taddia\inst{1}\and
I.~Arcavi\inst{3,4}\and
S.~B.~Cenko\inst{5,6}\and
K.~Clubb\inst{7}\and
A.~De~Cia\inst{8}\and
G.~Duggan\inst{2}\and
A.~V.~Filippenko\inst{7,9}\and
A.~Gal-Yam\inst{10}\and
M.~L.~Graham\inst{11}\and
A.~Horesh\inst{12}\and
G.~Hosseinzadeh\inst{4,3}\and
D.~A.~Howell\inst{4,3}\and
D.~Kuesters\inst{13}\and
R.~Lunnan\inst{1,2}\and
T.~Matheson\inst{14}\and
P.~E.~Nugent\inst{7,15}\and
D.~A.~Perley\inst{16}\and
R.~M.~Quimby\inst{17,18}\and
C.~Saunders\inst{19}}

\institute{Department of Astronomy, The Oskar Klein Center, Stockholm University, AlbaNova, 10691 Stockholm, Sweden\and
Department of Astronomy, California Institute of Technology, Pasadena, CA 91125, USA\and
Las Cumbres Observatory, 6740 Cortona Dr., Suite 102, Goleta, CA 93117, USA\and
Department of Physics, Broida Hall, University of California, Santa Barbara, CA 93106-9530, USA\and
NASA Goddard Space Flight Center, Mail Code 661, Greenbelt, MD 20771, USA\and
Joint Space-Science Institute, University of Maryland, College Park, MD 20742, USA \and
Department of Astronomy, University of California, Berkeley, CA 94720-3411, USA\and
European Southern Observatory, Karl-Schwarzschild-Strasse 2, 85748 Garching bei M\"unchen, Germany\and
Miller Senior Fellow, Miller Institute for Basic Research in Science, 
University of California, Berkeley, CA 94720, USA.\and
Benoziyo Center for Astrophysics, Weizmann Institute of Science, 76100 Rehovot, Israel\and
Astronomy Department, University of Washington, Box 351580, U.W., Seattle, WA 98195-1580, USA \and
Racah Institute of Physics, Hebrew University, Jerusalem, 91904, Israel\and
Institut f\"ur Physik, Humboldt-Universit\"at zu Berlin,
   Newtonstra\ss e 15, 12489 Berlin, Germany\and
National Optical Astronomy Observatory, 950 North Cherry Avenue, Tucson, AZ 85719, USA \and
Lawrence Berkeley National Laboratory, 1 Cyclotron Road, MS 50B-4206, Berkeley, CA 94720, USA\and
Astrophysics Research Institute, Liverpool John Moores University, IC2 Liverpool Science Park, 146 Brownlow Hill, Liverpool, L3 5RF, UK\and
Department of Astronomy, San Diego State University, San Diego, CA 92182, USA\and
9 Kavli IPMU (WPI), UTIAS, The University of Tokyo, Kashiwa, Chiba 277-8583, Japan\and
Laboratoire de Physique Nucl\'eaire et de Hautes \'Energies,
   Universit\'e Pierre et Marie Curie Paris 6, Universit\'e Paris Diderot Paris 7, CNRS-IN2P3,
   4 place Jussieu, 75252 Paris Cedex 05, France}

\date{Received; Accepted}

%\begin{abstract}
\abstract{
We present an analysis of 507 spectra of 173 stripped-envelope (SE) supernovae (SNe) discovered by the untargeted Palomar Transient Factory (PTF) and intermediate PTF (iPTF) surveys. Our sample contains 55 Type IIb SNe (SNe IIb), 45 Type Ib SNe (SNe Ib), 56 Type Ic SNe (SNe Ic), and 17 Type Ib/c SNe (SNe Ib/c). We compare the SE SN subtypes via measurements of the pseudo-equivalent widths (pEWs) and velocities of the \ion{He}{i}~$\lambda\lambda5876, 7065$ and \ion{O}{i}~$\lambda7774$ absorption lines. 
%Generally we observe a continuum in the properties among the three SE SN subtypes with no clear gaps in the distributions of individual measurements.
Consistent with previous work, we find that SNe Ic show higher pEWs and velocities in \ion{O}{i}~$\lambda7774$ compared to SNe IIb and Ib. % ($50$~\AA, and $\sim2000$~km~s$^{-1}$, respectively).
The pEWs of the \ion{He}{i}~$\lambda\lambda5876, 7065$ lines are similar in SNe Ib and IIb after maximum light. %with a few SNe Ib showing stronger absorptions prior to maximum light in the \ion{He}{i}~$\lambda$5876 line.
The \ion{He}{i}~$\lambda\lambda5876, 7065$ velocities at maximum light are higher % by $\sim1000$~km~s$^{-1}$
in SNe Ib compared to SNe IIb. We have identified an anticorrelation between the \ion{He}{i}~$\lambda7065$ pEW and \ion{O}{i}~$\lambda7774$ velocity among SNe IIb and Ib. This can be interpreted as a continuum in the amount of \ion{He}{} present at the time of explosion. It has been suggested that SNe Ib and Ic have similar amounts of He, and that lower mixing could be responsible for hiding He in SNe Ic. However, our data contradict this mixing hypothesis. %\textbf{Higher velocities in SNe Ic could be explained by higher kinetic energies, but this would be inconsistent with the anticorrelation between \ion{He}{i} and \ion{O}{i}, as higher kinetic energy would not be expected to systematically decrease the mixing and the \ion{He}{i} absorption strength.} %The anticorrelation between \ion{He}{i} absorption strength and \ion{O}{i} line velocity for SNe Ib and IIb can be interpreted as a difference in the amount of \ion{He}{} present at the time of explosion. %%This could be on account of a difference in starting He core masses or due to a continuum in the amount of stripping experienced by the progenitor stars, or a combination thereof.
The observed difference in the expansion rate of the ejecta around maximum light of SNe Ic ($V_{\mathrm{m}}=\sqrt{2E_{\mathrm{k}}/M_{\mathrm{ej}}}\approx15,000$~km~s$^{-1}$) and SNe Ib ($V_{\mathrm{m}}\approx9000$~km~s$^{-1}$) would imply an average He mass difference of $\sim1.4$~\msun, if the other explosion parameters are assumed to be unchanged between the SE SN subtypes. We conclude that SNe Ic do not hide He but lose He due to envelope stripping.}
%\end{abstract}

\maketitle

%\titlerunning{Oxygen and helium in stripped-envelope supernovae}
%\authorrunning{C.~Fremling et al.}

\section{Introduction}\label{sec:intro}

Stripped-envelope (SE) supernovae (SNe) are thought to be the result of massive stars undergoing core collapse (CC) after the progenitor stars have been stripped of their envelopes to varying degrees. Among SE SNe there are three main subtypes: Type IIb SNe (SNe IIb), Type Ib SNe (SNe Ib) and Type Ic SNe (SNe Ic). Observationally, the distinction between the subtypes is based on the presence of \ion{H}{} and \ion{He}{} lines. SNe IIb show helium and hydrogen signatures in early-time spectra, with the hydrogen signatures disappearing over time. SNe Ib show no hydrogen but strong helium features, and SNe Ic show neither hydrogen nor helium lines in their spectra \citep{1997ARA&amp;A..35..309F,2016arXiv161109353G}. The spectral differences among the SE SN subtypes are typically interpreted as varying amounts of envelope stripping. SNe IIb progenitors would then have undergone partial stripping, retaining a small part of their hydrogen envelopes. SNe Ib progenitors would be fully stripped of their hydrogen envelopes, and SNe Ic progenitors would be fully stripped of both their hydrogen and helium envelopes.

There are two main mechanisms that can give rise to significant envelope stripping: line-driven winds from isolated massive stars (e.g., \citealp{conti76}; \citealp{2014ARA&amp;A..52..487S}) or binary mass transfer (e.g., \citealp{1985ApJS...58..661I}; \citealp{Yoon:2010aa}; \citealp{2011A&amp;A...528A.131C}; \citealp{2015PASA...32...15Y}). However, alternative explanations for the observed subtypes have been suggested. 
Enhanced stellar mixing in massive stars could reduce the \ion{He}
envelope mass by burning it into \ion{O}, which produces SNe Ic 
with enhanced \ion{O}{} abundances (\citealp{2013ApJ...773L...7F}), and differences in the mixing of
\Nif\ synthesized in the SN explosions could produce SNe of different
observational subtypes from progenitors of otherwise similar structure
(\citealp{2012MNRAS.424.2139D}). In the models by
\cite{2012MNRAS.424.2139D}, only SN explosions where the synthesized
\Nif\ is highly mixed throughout the ejecta, so that the \ion{He}{} in the envelope can be nonthermally excited, will produce detectable \ion{He}{} lines, as seen in SNe Ib. Models with low mixing become SNe Ic. In these models, higher mixing also results in stronger and faster \ion{O}{i}~$\lambda7774$\footnote{When discussing this line we effectively refer to the \ion{O}{i}~$\lambda\lambda7771, 7774, 7775$ triplet.}, and thus SNe Ib would show the strongest and fastest oxygen (see fig.~13 in \citealp{2012MNRAS.424.2139D}). 

Observationally, there is some evidence both for and against these
scenarios. \cite{2014ApJ...792L..11P} suggest that up to $1$~\Msun\ of
\ion{He}{} could sometimes be hidden in SE SNe, as some objects 
show very low \ion{He}{} velocities indicating that the emitting region
lies behind a transparent shell (e.g., the Type IIb
SN~2010as; \citealp{2014ApJ...792....7F}). In such objects, based on
the calculations by \cite{2012MNRAS.424.2139D}, the \Nif\ synthesized
in the explosion might only be mixed into a small part of the \ion{He}
envelope, resulting in lower velocities measured from absorption
minima, and indicating the possibility that even lower mixing, if
possible, could produce no \ion{He}{} signatures at all (and thus give
rise to SNe Ic with very slow expansion velocities). In contrast
to this suggestion, it has been found that SNe Ic 
on average display the highest velocities, 
followed by SNe Ib, and finally by SNe IIb at the slowest expansion velocities as measured in both the \ion{Fe}{ii}~$\lambda5169$ and the \ion{O}{i}~$\lambda7774$ absorption lines \citep{2001AJ....121.1648M,2016ApJ...827...90L}. However, these results were based on a relatively low number of objects, especially for the \ion{O}{i}~$\lambda7774$ line. The SNe included in these studies were also discovered mainly in targeted SN searches. %\cite{2017MNRAS.469.2672P} have also performed an analysis on the same sample as \cite{2016ApJ...827...90L} with some additional SNe, but this was focused on the \ion{H\alpha} line and the \ion{He}{} lines.

In this paper, we perform an analysis similar to that of \cite{2016ApJ...827...90L}, in order to check if their results hold with a larger sample. We investigate \ion{He}{i}~$\lambda\lambda5876, 7065$ and
\ion{O}{i}~$\lambda7774$ absorption-line strengths and velocities
during the photospheric phase for all 176 SE SNe discovered by the
Palomar Transient Factory (PTF; \citealp{Law:2009aa}) and the
intermediate PTF (iPTF; \citealp{2016PASP..128k4502C,2017PASP..129a4002M}). The PTF and iPTF were untargeted
magnitude-limited\footnote{Approximately 20.5 and 21 mag in the Mould $R$ and Sloan
  $g$ bands, respectively.} surveys. Thus, our sample should arguably be
less biased compared to previous studies of SE SN spectra that have been mostly based on SNe found via targeted searches\footnote{This is  potentially important since (i)PTF could find SE SNe in hosts with very low metallicity compared to the average for SE SNe (e.g., \citealp{2012ApJ...758..132S}). Two examples are iPTF15dtg \citep{2016A&amp;A...592A..89T} and PTF11mnb \citep{2017arXiv170908386T}. However, their spectra are similar to those of normal SNe~Ic, and we do not find any significant differences in our results compared to \cite{2001AJ....121.1648M} and \cite{2016ApJ...827...90L}. Thus, a detailed analysis of the impact of (i)PTF being untargeted was not performed.}. We address
the predictions of alternative models for the observed subtypes of SE
SNe\footnote{By comparing observed spectra to the models by \cite{2012MNRAS.424.2139D}, as was previously done by \cite{2016ApJ...827...90L}.}, and compare our results to those found by \cite{2001AJ....121.1648M}
and \cite{2016ApJ...827...90L}. %Our investigation differs somewhat from
%that of \cite{2016ApJ...827...90L}; we attempt measurements of
%all lines, regardless of the classification. 
Our observations are described in Sect.~\ref{sec:observations}. The methods we have used are presented in Sect.~\ref{sec:analysis}, and the results are given in Sect.~\ref{sec:results}. We search for correlations between the various measurements in Sect.~\ref{sec:correlations}. A discussion and our conclusions can be found in Sect.~\ref{sec:discussion}. 

\section{Observations and reductions}
\label{sec:observations}
\subsection{Photometry}
To estimate the epoch of maximum light for each SN, we utilize $r$- or $g$-band photometry from the Palomar 48~inch (P48) and 60~inch (P60) telescopes, depending on availability for each SN. All photometry has been host-galaxy subtracted. The P60 data were reduced with {\sc FPipe} \citep{2016A&amp;A...593A..68F}, and the P48 data with the {\sc PTFIDE} pipeline \citep{2017PASP..129a4002M}. The full photometric (i)PTF SE SN sample will be presented in future papers. A subsample has previously been analyzed by \cite{2016MNRAS.458.2973P}.%by Karamehmetoglu et al. (in prep.) and Barbarino et al. (in prep.).%LCOGT data was reduced with the Valenti pipeline \citep{2016MNRAS.459.3939V}. 

\subsection{Spectroscopy}
During the lifetimes of PTF and iPTF we obtained $507$ spectra at
unique epochs of SE SNe. We discovered 55 SNe IIb (187 spectra), 45 SNe Ib (125 spectra), 56 SNe Ic (153 spectra), and 17 Type Ib/c SNe (SNe Ib/c; 42 spectra). We have over three times more objects in our sample compared to \cite{2016ApJ...827...90L}, who studied 14 SNe IIb, 21 SNe Ib, and 17 SNe Ic. The (i)PTF SE SN spectral sample is
summarized in Table~\ref{tab:spec0}.

We do not
include Type~Ic-BL SNe discovered by the (i)PTF in the analysis in
this paper; they will be presented by Taddia et al. (in
prep.). Hydrogen-poor superluminous SNe (SLSNe) are also excluded. %, except for borderline cases (see Table~\ref{tab:spec0}). SLSNe from (i)PTF are being investigated by Quimby et al., (in prep.).
Subsamples of our full spectral dataset have previously been
published: some SNe IIb by \cite{2015ApJ...811..117S}, iPTF13bvn and PTF12os by \cite{2014A&amp;A...565A.114F,2016A&amp;A...593A..68F}, iPTF15dtg by \cite{2016A&amp;A...592A..89T}, SN~2013cu (iPTF13ast) by \cite{2014Natur.509..471G}, PTF12gzk by \cite{2013ApJ...778...63H}, SN~2011dh (PTF11eon) by \cite{2011ApJ...742L..18A}, and SN~2010mb (PTF10iue) by \cite{2014ApJ...785...37B}.
%and a selection of Type Ic-BL by \citealp{2016ApJ...830...42C}

All spectra included in our analysis were reduced using standard pipelines and procedures for each telescope and instrument. The analysis in this paper is focused on the optically thick photospheric phase (we do not analyze spectra taken later than 60~d past maximum light). Normalized spectra limited to the regions around the absorption features used in our analysis will be made available in electronic form via WISeREP\footnote{\href{http://www.weizmann.ac.il/astrophysics/wiserep/}{http://www.weizmann.ac.il/astrophysics/wiserep/}} \citep{Yaron:2012aa} and the Open Supernova Catalog\footnote{\href{https://sne.space}{https://sne.space}} (OSC; \citealp{2017ApJ...835...64G}). %All spectral data and corresponding information will be made available via WISeREP \footnote{\href{http://www.weizmann.ac.il/astrophysics/wiserep/}{http://www.weizmann.ac.il/astrophysics/wiserep/}} \citep{Yaron:2012aa}.
A catalogue paper that will present full spectra %that will present all of our spectra obtained with the Keck, the Palomar 200-inch, the Nordic Optical Telescope (NOT), the Lick, and the Gemini telescopes
is in preparation.

%%% also make available on open supernova catalog 

\section{Data analysis and methods}\label{sec:analysis}

\subsection{Time of Maximum Light Estimates}
We estimate the time of maximum light for each SN in our sample using the observed light curves (LCs), by performing template fits to the $r$-band LC of each object (or the $g$-band LC if $r$ is not available). When fitting the LCs we allow for a shift and stretch of the templates. The LC peaks of SNe IIb were estimated using the $r$-band LC of SN~2011dh \citep{2014A&amp;A...562A..17E}. The LC peaks of SNe Ib and Ic were estimated using the LC templates by \cite{2015A&amp;A...574A..60T}.

The time of maximum light was used to calculate the rest-frame epochs of our spectroscopic observations (see Table~\ref{tab:spec0}). Throughout this paper we use the convention that negative epochs refer to epochs before maximum light (e.g., $-5$~d) and positive epochs refer to past maximum (e.g., $+5$~d). For objects where it was not possible to determine the time of maximum light, we use the spectra obtained by the (i)PTF to classify them (see Table~\ref{tab:spec0}), but do not include them in any further analysis.

\subsection{Spectral Classifications}
To classify the SNe in our sample we have used {\sc SNID} (\citealp{2007ApJ...666.1024B}) with template spectra constructed from the SE SNe in the Harvard-Smithsonian Center for Astrophysics sample \citep{2014AJ....147...99M,2014arXiv1405.1437L}. The full spectral sequence for each SN has been run automatically through {\sc SNID}, and the subtype with the most matches across the spectral sequence (before +60~d) is used to classify the SN.

If there are several conflicting matches for different spectra of the same SN, or if there is a similar number of matches for two subtypes for a certain spectral sequence, the results are manually inspected both in {\sc SNID} and by inspecting the regions around \ion{H}{}$\alpha$, \ion{H}{}$\beta$, and \ion{He}{i}~$\lambda\lambda5876, 7065$; if the best {\sc SNID} matches change over time from SNe IIb to SNe Ib, or if both \ion{H}{}$\alpha$ and \ion{H}{}$\beta$ absorption is detected at a similar velocity in any early-time spectrum, we classify the object as a SN IIb. Furthermore, if manual inspection is needed, \ion{H}{} is undetected, and absorption from both of the \ion{He}{i}~$\lambda\lambda5876, 7065$ lines is detected, we classify the object as a SN Ib; if neither \ion{H}{} nor \ion{He}{} is detected, the object is classified as a SN Ic (see also \citealp{2017PASP..129e4201S}, who also discuss the difficulty of classifying some SE SNe).

SNe with spectra where the signal-to-noise ratio (SNR) is not sufficiently good to distinguish a SN Ib from a SN Ic, or where the classification remains ambiguous after manual inspection, are classified as SNe Ib/c. These are not included in our analysis, except to check for the impact on average values (Sect.~\ref{sec:results}) and correlations (Sect.~\ref{sec:correlations}) by including these objects in both the SN Ib and SN Ic groups and redoing the calculations. We find no significant changes in any of our conclusions by doing this exercise.

\subsection{Pseudo-Equivalent Widths}\label{sec:pewmethod}
To estimate the strength of absorption features in our spectra we utilize the pseudo-equivalent width (pEW; e.g., \citealp{2011A&amp;A...526A.119N}), defined as
\begin{equation}\label{eq:pew}
\mathrm{pEW}=\sum_{i=1}^{N}\left({1-\frac{f(\lambda_i)}{f_{0}(\lambda_i)} }\right)\Delta\lambda_i,
\end{equation}
where $\lambda_1$ and $\lambda_N$ define the estimated extent of the absorption feature ($\lambda_1<\lambda_N$), and $f_0(\lambda_i)$ is the continuum, estimated as a linear fit to the data surrounding $\lambda_1$ and $\lambda_N$. 

\begin{figure*}[ht]
\centering
\includegraphics[width=16cm]{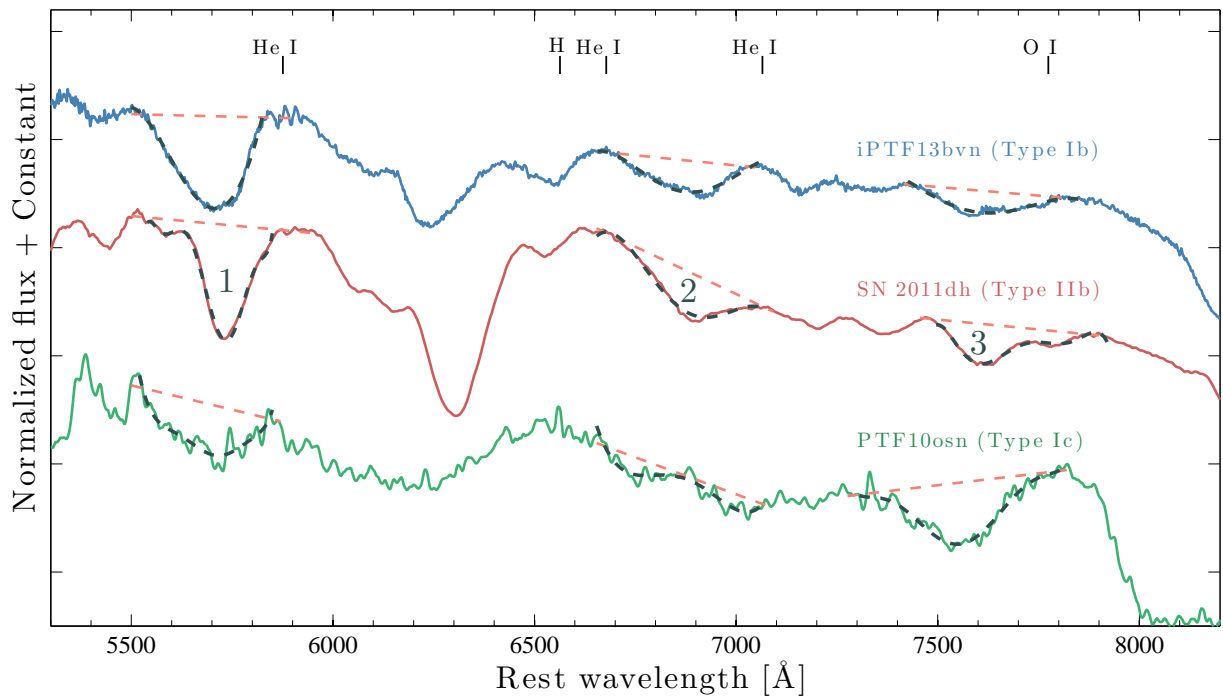}
\caption{Description of our pEW and expansion-velocity measurement method using spectra around peak brightness for iPTF13bvn (Type Ib, +0.5~d, upper blue line), SN~2011dh (Type IIb, +2.5~d, middle red line), and PTF10osn (Type Ic, +4.0~d, bottom green line). On the spectrum of SN~2011dh, the absorption features of \ion{He}{i}~$\lambda\lambda5876, 7065$ and \ion{O}{i}~$\lambda$7774 have been labeled with numbers. Measurements of features 1-3 have been attempted on all spectra in our sample. For SE SNe, feature 1 is typically associated with \ion{He}{i}~$\lambda$5876, feature 2 with \ion{He}{i}~$\lambda$7065, and feature 3 with \ion{O}{i}~$\lambda$7774, although other elements also contribute to these features. Dashed red lines indicate pseudo-continuum estimates (first-order polynomial fits) for features 1-3 resulting from our measurement scheme (Sect.~\ref{sec:analysis}). We calculate pEWs (e.g., Fig.~\ref{fig:pew_hei}) between the endpoints of these continuum estimates using Eq.~\ref{eq:pew}. Velocity estimates are obtained by locating the minima of polynomial fits between the pseudo-continuum endpoints (order 4-7, dashed black lines). If a spectral feature is ambiguous, and has several minima within a similar range as shown for feature 2 in the spectrum of PTF10osn, we set the velocity estimate to be undefined but still calculate the pEW. Wavelengths have been shifted to the rest frame, and the spectra have been normalized and shifted by a constant for clarity.}
\label{fig:spec_sequence_method}
\end{figure*}

We follow similar principles as \cite{2016ApJ...827...90L} when
choosing the endpoints of the absorption features and where to measure
the continuum levels. First, we smooth the spectrum\footnote{Final
  measurements are performed on unsmoothed spectra. Only $\lambda_1$ and $\lambda_N$ are located using smoothed spectra.}
around the absorption feature. The first peak on the blue side of the
feature is then chosen as $\lambda_1$ and the peak closest to the
expected position for the relevant emission-line peak in the rest
frame is chosen as $\lambda_N$. Typical results of this method are
shown for an example spectrum taken close to maximum light for each SE
SN subtype in Fig.~\ref{fig:spec_sequence_method} (we show Type Ib
iPTF13bvn, data from \citealp{2014A&amp;A...565A.114F};
Type IIb SN~2011dh\footnote{See \cite{2014A&amp;A...562A..17E,
  2015A&amp;A...580A.142E}, for a detailed investigation of SN~2011dh.}, data from this work; and Type Ic PTF10osn, data from this work). Note
that if there are multiple peaks within $\sim100$~\AA\ of the apparent
ends of the absorption feature, we choose the peaks that give the
highest pEW value as $\lambda_1$ and $\lambda_N$ (see feature 1 in the
spectrum of SN~2011dh in Fig.~\ref{fig:spec_sequence_method}). In the
case where there is no clear peak on the blue side of the absorption
feature, we limit the pEW measurement in velocity space to no larger
than $25,000$~km~s$^{-1}$ for the position of $\lambda_1$. If there is
no clear absorption feature and no clear emission peak to use as
$\lambda_N$ (typical especially for the \ion{He}{i} $\lambda$7065
position in SNe Ic), we fix $\lambda_N$ at the expected rest-frame
position of the emission feature and set $\lambda_1$ where the value
of pEW is maximized while $\lambda_1$ is below the maximum allowed
velocity (see feature 2 in the spectrum of PTF10osn in Fig.~\ref{fig:spec_sequence_method}).

Uncertainties in the pEW measurements are estimated using a Monte-Carlo method. First, we create a local measurement of the noise by dividing the original spectrum with a heavily smoothed spectrum and computing the standard deviation locally (in a region of about $1000$~\AA) around the position of the absorption feature to be measured. Next, we create many simulated noisy spectra by adding noise using a Gaussian distribution and the standard deviation. The pEW measurement is then repeated on each simulated spectrum and the standard deviation of the results is taken as the 1$\sigma$ uncertainty of our measurements. In this procedure, we also randomly change the continuum endpoints in each simulated spectrum within a $25$~\AA\ region to account for possible uncertainties in the identifications of the peak positions of the spectral features. Although this is not necessary for good-quality spectra with accurate peak position estimates, we still perform the same calculation regardless of the spectral quality for consistency\footnote{For spectra of very low SNR we get uncertainties on the order of $25$~\AA\ in our peak position estimates.}. 

\subsection{Expansion Velocities}
To estimate expansion velocities we use the minima of the identified absorption features in our pEW measurements. This is done by fitting a polynomial to the absorption feature and locating the minimum of the fit\footnote{We use polynomials of degree 4 to 7 depending on the feature shape and quality of the spectrum.}. Example fits are shown in Fig.~\ref{fig:spec_sequence_method} (dashed black lines). Uncertainties are estimated using a similar MC approach as for the pEW measurements. The minimum of each simulated spectrum is estimated by polynomial fits, and the standard deviation is taken as the 1$\sigma$ uncertainty of our velocity measurements.

\section{Results}\label{sec:results}
\subsection{Absorption Strengths}\label{sec:pew}
The method we use to measure pEWs (Sect.~\ref{sec:pewmethod}) will by
construction tend to give positive values for features 1 and 2 as
identified in Fig.~\ref{fig:spec_sequence_method} for all SE SN
spectra, including those of SNe Ic. In SNe IIb and Ib,
these features are typically identified as \ion{He}{i}~$\lambda\lambda5876, 7065$. However, positive pEW measurements in a
SN Ic does not 
necessarily
mean that \ion{He}{} is present in the ejecta. 
Our method simply maximizes the pEW measurement of any absorption from any line that could give rise to a feature in these regions of the spectrum. Possible contamination for the \ion{He}{i} $\lambda$5876 absorption line is \ion{Na}{i\,D}, and for \ion{He}{i} $\lambda$7065, \ion{Al}{ii} (e.g., \citealp{2010ApJ...723L..98K}). Regardless, throughout this paper we will refer to any measurable absorption found at the position of features 1 and 2 as potential \ion{He}{i}~$\lambda\lambda5876, 7065$ absorption, even for SNe Ic, to simplify the discussion.

\subsubsection{Helium}
\begin{figure*}[t]
\centering
\vspace{-0.2cm}
\includegraphics[width=7.8cm]{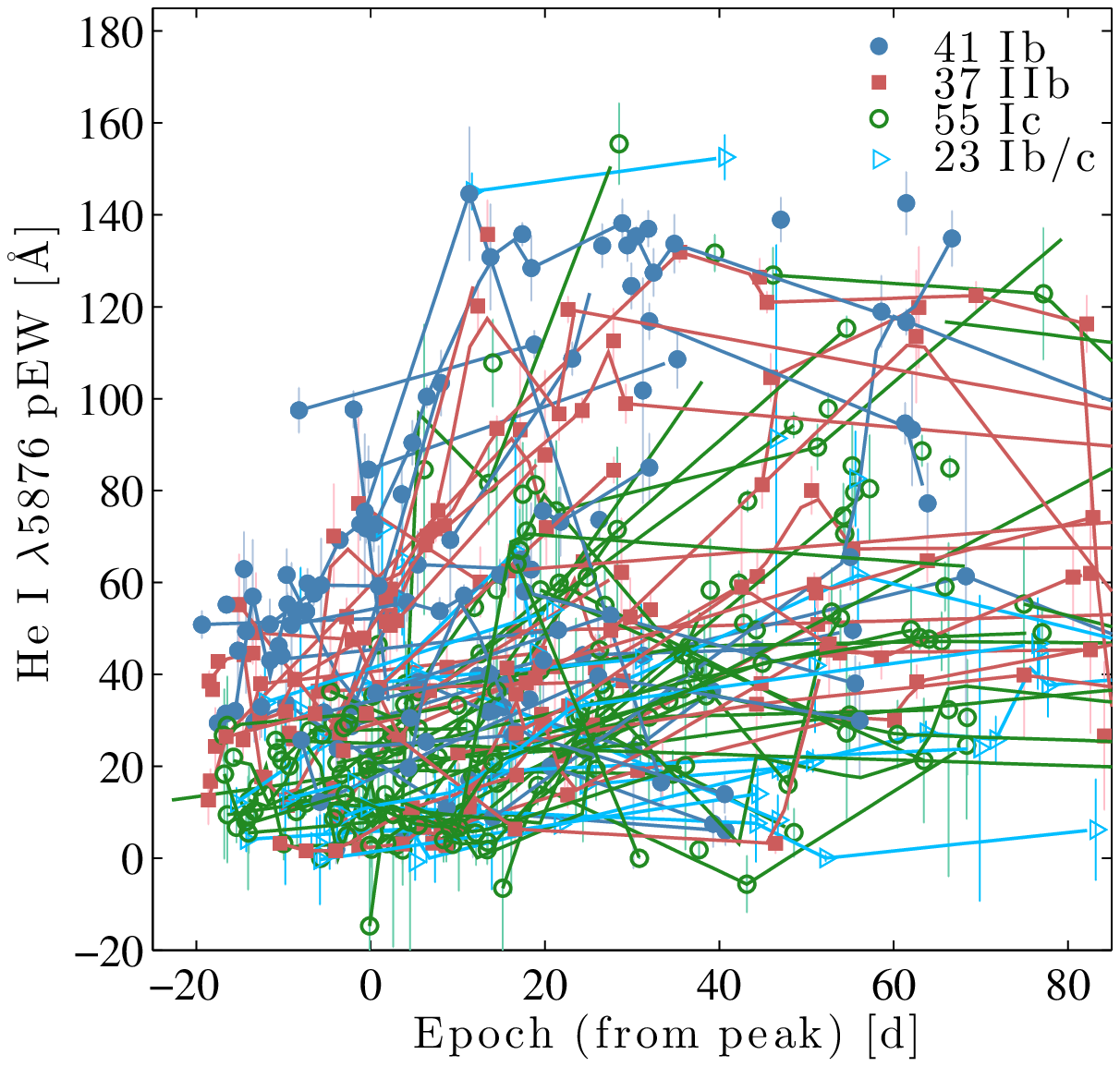}\includegraphics[width=7.85cm,trim=0 0.07cm 0 0]{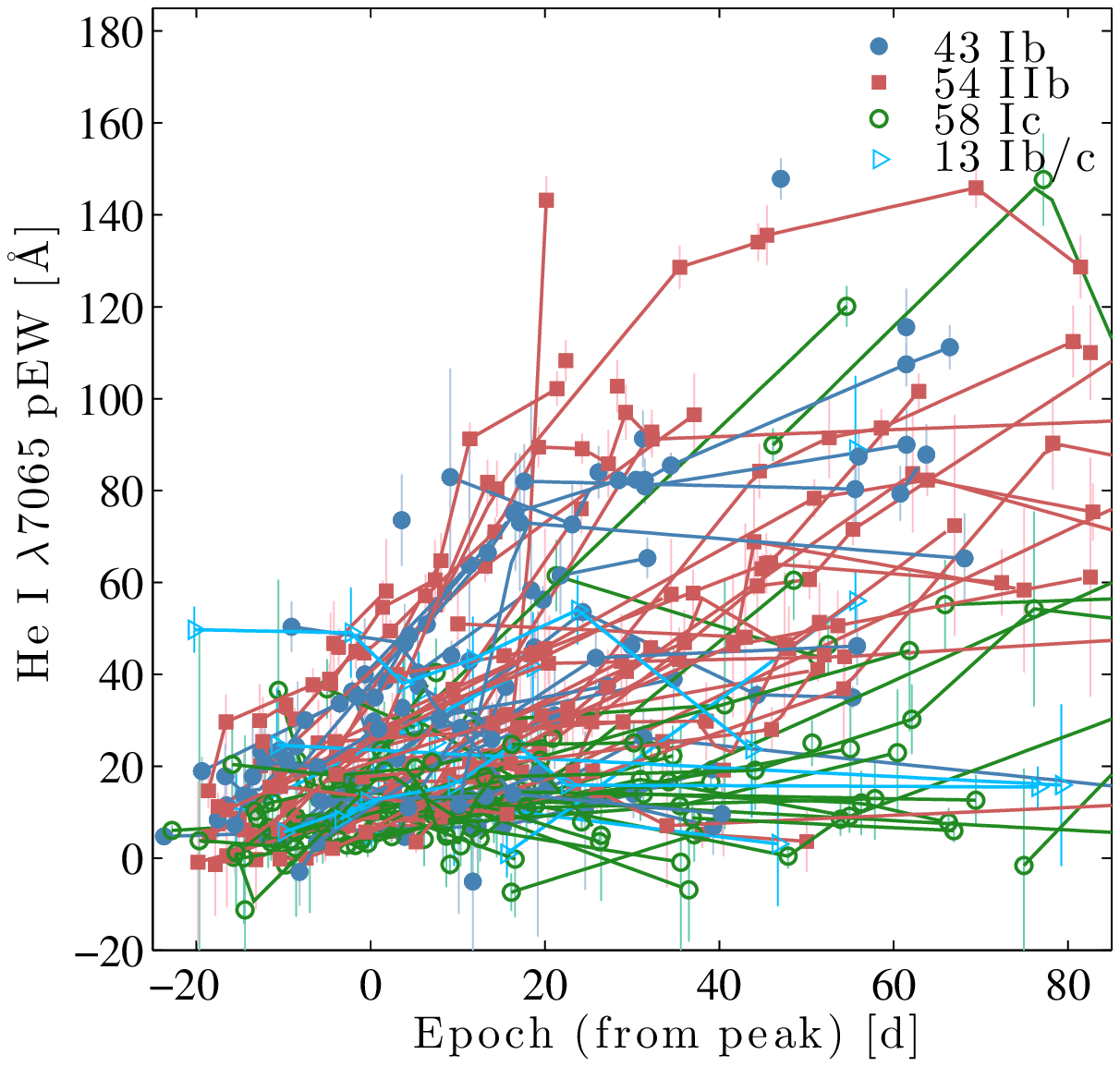}
\\
\includegraphics[width=7.8cm]{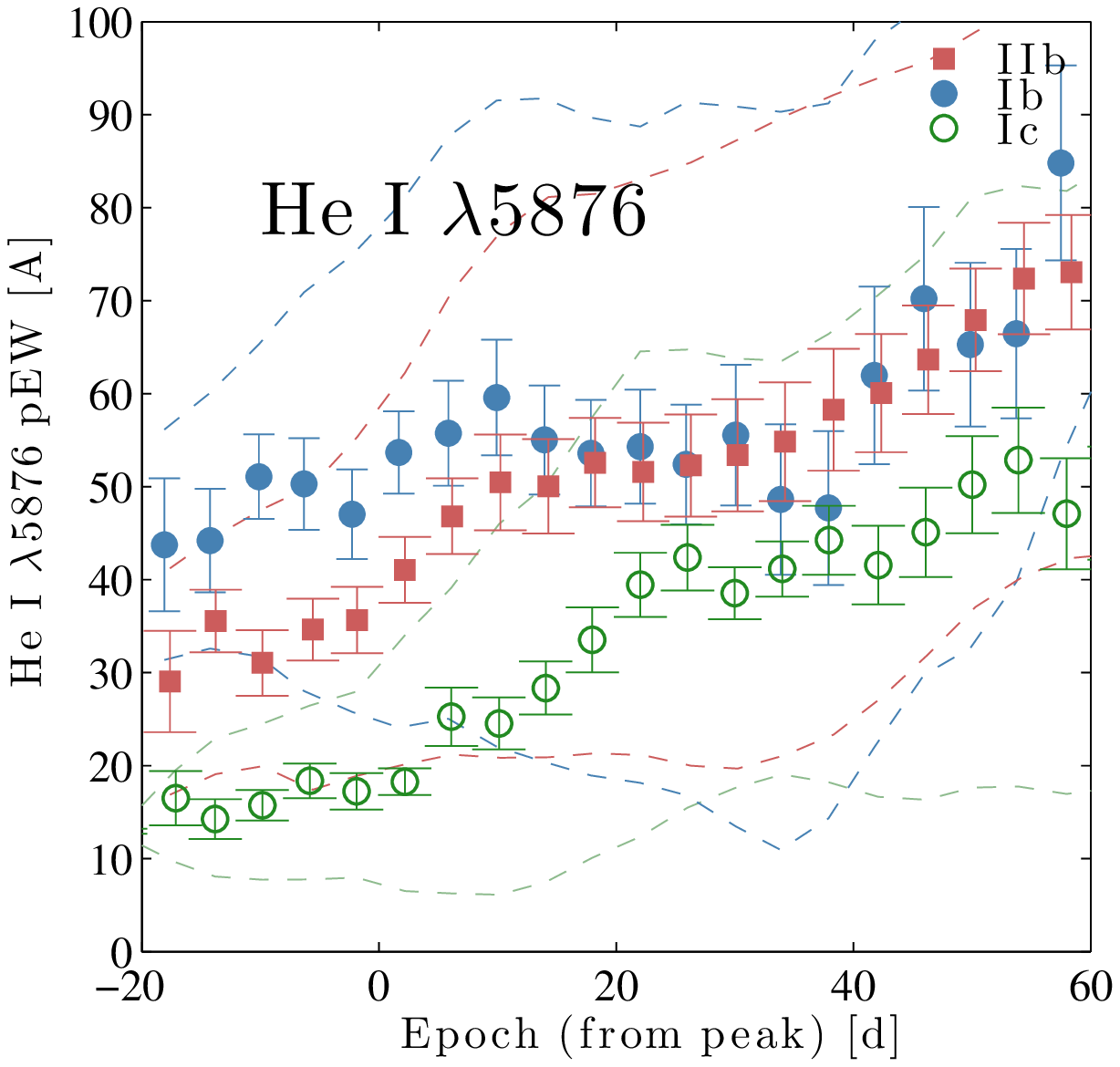}\includegraphics[width=7.88cm,trim=0 0.1cm 0 0]{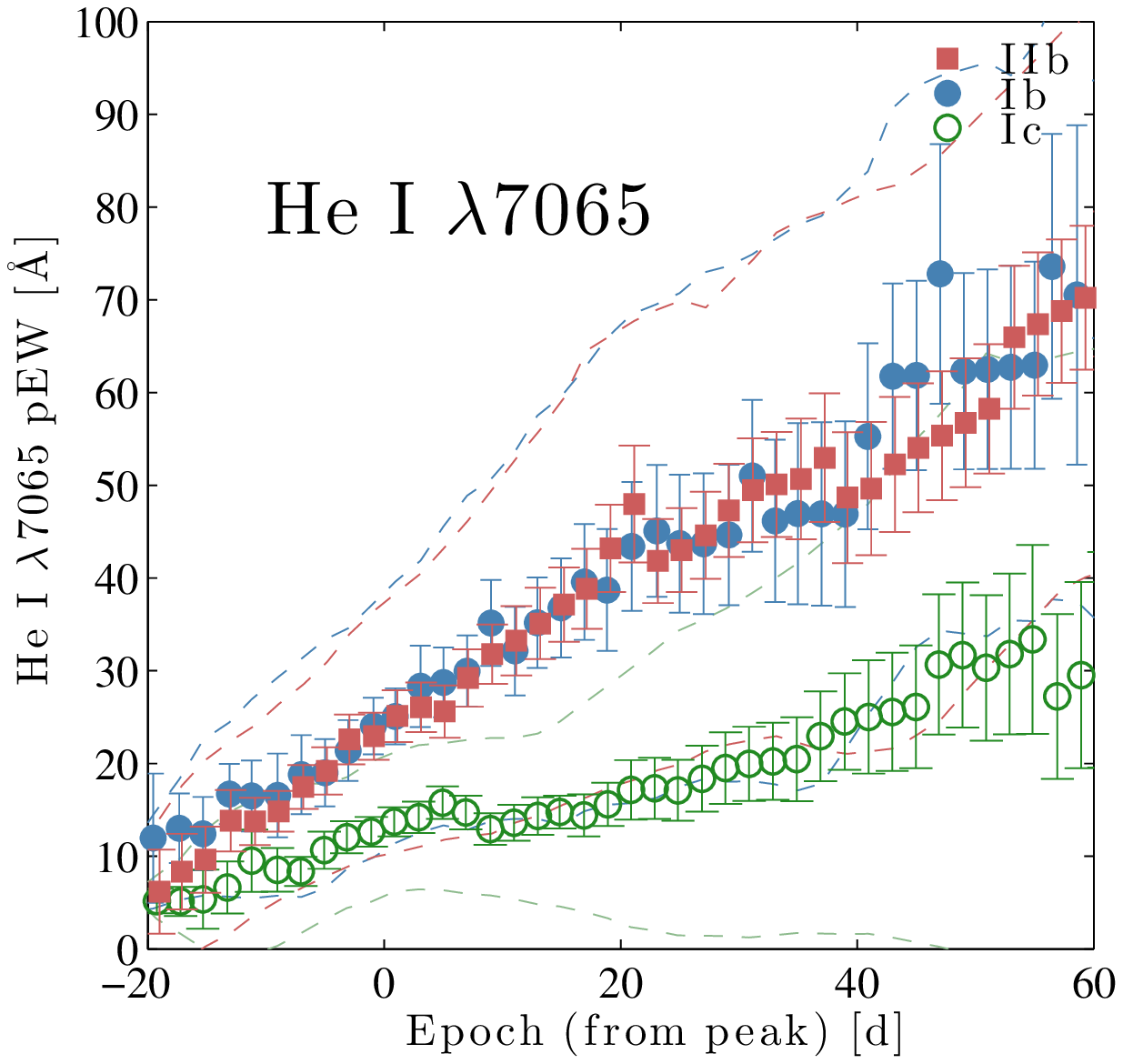}
\\
\includegraphics[width=7.8cm]{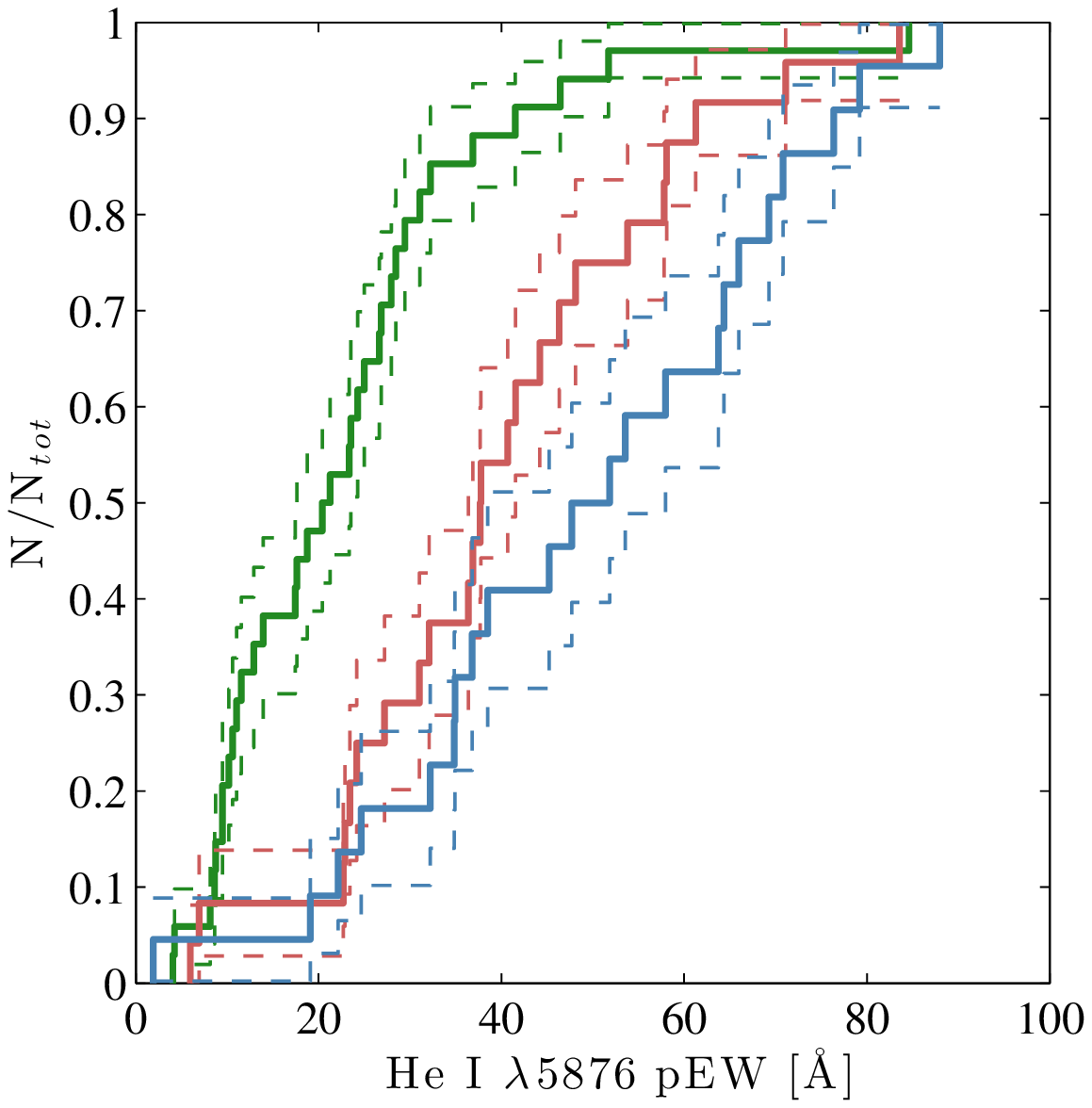}\includegraphics[width=7.85cm,trim=0 0.1cm 0 0]{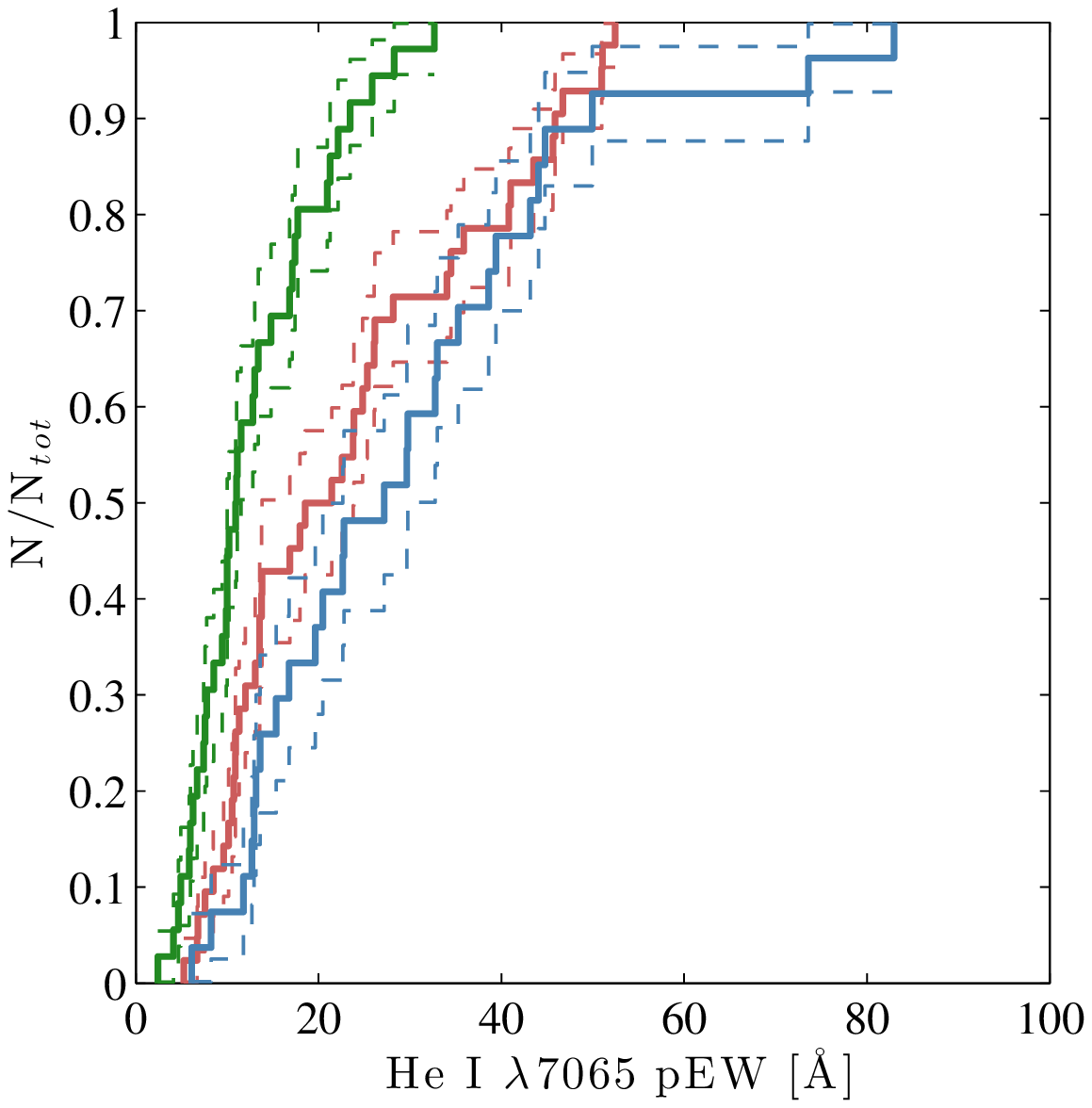}
\caption{Absorption strengths of \ion{He}{i}~$\lambda$5876 (left panels) and \ion{He}{i}~$\lambda$7065 (right panels). The top panels show individual measurements, with multiple measurements of the same SN connected by solid lines. The middle panels show averages of the SE SN subtypes, with error bars representing the standard deviation of the mean. Dashed lines outline the standard deviations of the samples for each subtype in matching color. The bottom panels show CDFs measured between $-10$ and $+10$~d}
\label{fig:pew_hei}
\end{figure*}

We show the individual \ion{He}{i} pEW measurements of the SNe IIb, Ib, and Ic included in our sample in the top panels of Fig.~\ref{fig:pew_hei}. The evolution of the mean values and the standard deviation of the means of each subtype is shown in the middle panels of the figure. 

Similar to \cite{2016ApJ...827...90L}, we find that SNe Ib show larger pEW values on average compared to SNe IIb for the \ion{He}{i}~$\lambda5876$ feature before $+20$~d, after which the pEW values become similar. The probability that this difference is not real based on a two-sample Kolmogorov-Smirnov (K-S) test is $p<0.03$, when we consider the measurements between $-10$ to $+10$ days of each subtype. %\footnote{Similar $p$ values can be found for all time-intervals between $-20$ to $+10$ days. At later times there is no difference between the Type IIb and Type Ib SN distributions.}.
For the same time interval we show the cumulative distribution functions (CDFs) of the pEW values for each subtype in the bottom-left panel of Fig.~\ref{fig:pew_hei}. SNe Ic show significantly weaker absorption in this line at early times ($p<0.001$, before $+20$~d when compared to both our SNe IIb and Ib). However, the pEW values of SNe Ic become more similar to those of the other SE SNe after $+30$~d. This is likely a result of contamination from \ion{Na}{i} gradually becoming more significant, although a weak contribution from \ion{He}{i} cannot be ruled out.% (but see Sect.~\ref{sec:corr_helium} and Fig.~\ref{fig:pew_correlations_1}). 

In contrast to \cite{2016ApJ...827...90L}, we do not find a strong difference between SNe IIb and Ib for the pEW values of the \ion{He}{i}~$\lambda7065$ absorption line at epochs past $\sim+40$~d (see the middle panels of our Fig.~\ref{fig:pew_hei} and compare to fig.~6 in \citealp{2016ApJ...827...90L}). Instead, we find that SNe IIb and Ib are remarkably similar during all epochs when we could obtain meaningful averages based on measurements of multiple objects (see the bottom-right panel of Fig.~\ref{fig:pew_hei} for the CDFs for the $-10$~d to $+10$~d time interval). In a K-S test we do not find any statistically significant difference among our SNe IIb and Ib in \ion{He}{i}~$\lambda7065$ at any time. Our sample contains several SNe Ib that evolve toward much stronger absorption in this line compared to those studied by  \cite{2016ApJ...827...90L}. We also have several SNe IIb showing weaker absorption than those studied by \cite{2016ApJ...827...90L}. This could be due to the fact that the (i)PTF is untargeted, resulting in a greater diversity of objects, or a consequence of our larger sample size. %These could possibly be Type IIb SNe of the compact variety. It is also possible that they are less efficiently mixed compared to the average, causing weaker \ion{He}{i} lines (see \citealp{2012MNRAS.424.2139D}).

SNe Ic show consistently much weaker pEW values measured at the expected position for \ion{He}{i}~$\lambda7065$ absorption at all times, although we do observe a rising trend over time in the pEW values that is similar to that of the other subtypes. This could be evidence for a small residual \ion{He}{} envelope in some SNe Ic (but see also Sect.~\ref{sec:corr_helium} and Fig.~\ref{fig:pew_correlations_1}). It could also be due to contamination from the nearby [\ion{Fe}{i}]~$\lambda7155$ line that is gradually getting more significant, or to \ion{Al}{ii} absorption.

\begin{table}
\begin{adjustbox}{width=\columnwidth}
\begin{threeparttable}
\caption{\small{Weighted mean values for SE SNe at maximum light.}}
\scriptsize
\begin{tabular}{lccccccc}
\toprule
SN Type & pEW (\ion{O}{i}~$\lambda$7774) & $v$ (\ion{He}{i}~$\lambda$5876) & $v$ (\ion{O}{i}~$\lambda$7774) &\\ 
 & (\AA) & (km s$^{-1}$) & (km s$^{-1}$) \\
\midrule
SNe Ic &   70 $\pm$ 15 &       ... &  9800 $\pm$ 600\\
SNe Ib  &   28 $\pm$ 12 &       9500 $\pm$ 600 & 7900 $\pm$ 600\\
SNe IIb  &   32 $\pm$ 10 &       8000 $\pm$ 500 & 7000 $\pm$ 500 \\
\bottomrule
\end{tabular}
  \begin{tablenotes}
    \small
      \item \scriptsize{Note. --- Uncertainties are given as the standard deviation of the mean values, $\sigma/\sqrt{N}$, where $N$ is the number of objects included in the calculation.}
   \end{tablenotes}
\label{table:avgs}
\end{threeparttable}
\end{adjustbox}
\end{table}

\subsubsection{Oxygen}\label{sec:oxygen}
\begin{figure*}[t]
\centering
\includegraphics[width=8cm]{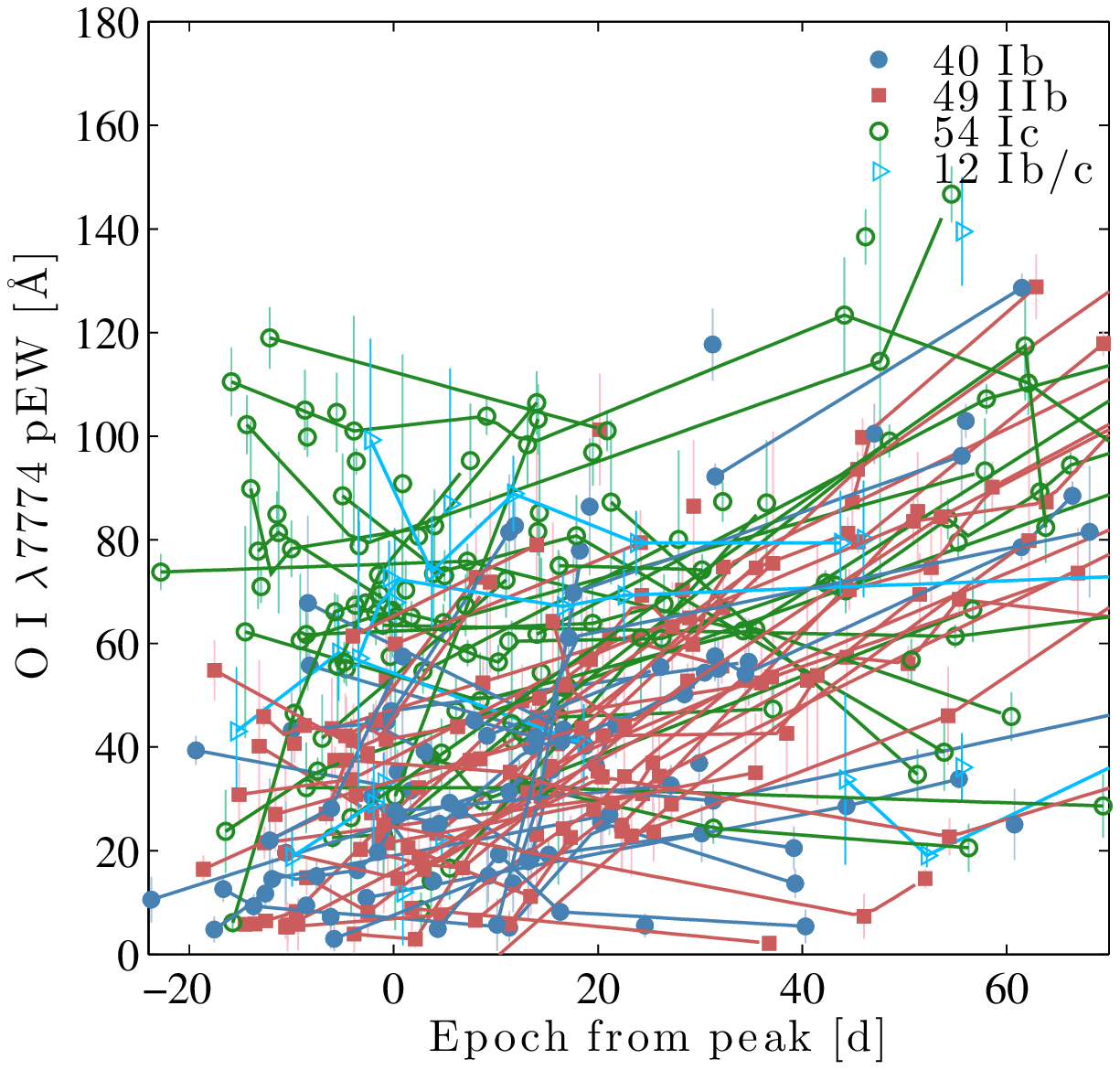}
\includegraphics[width=8.1cm]{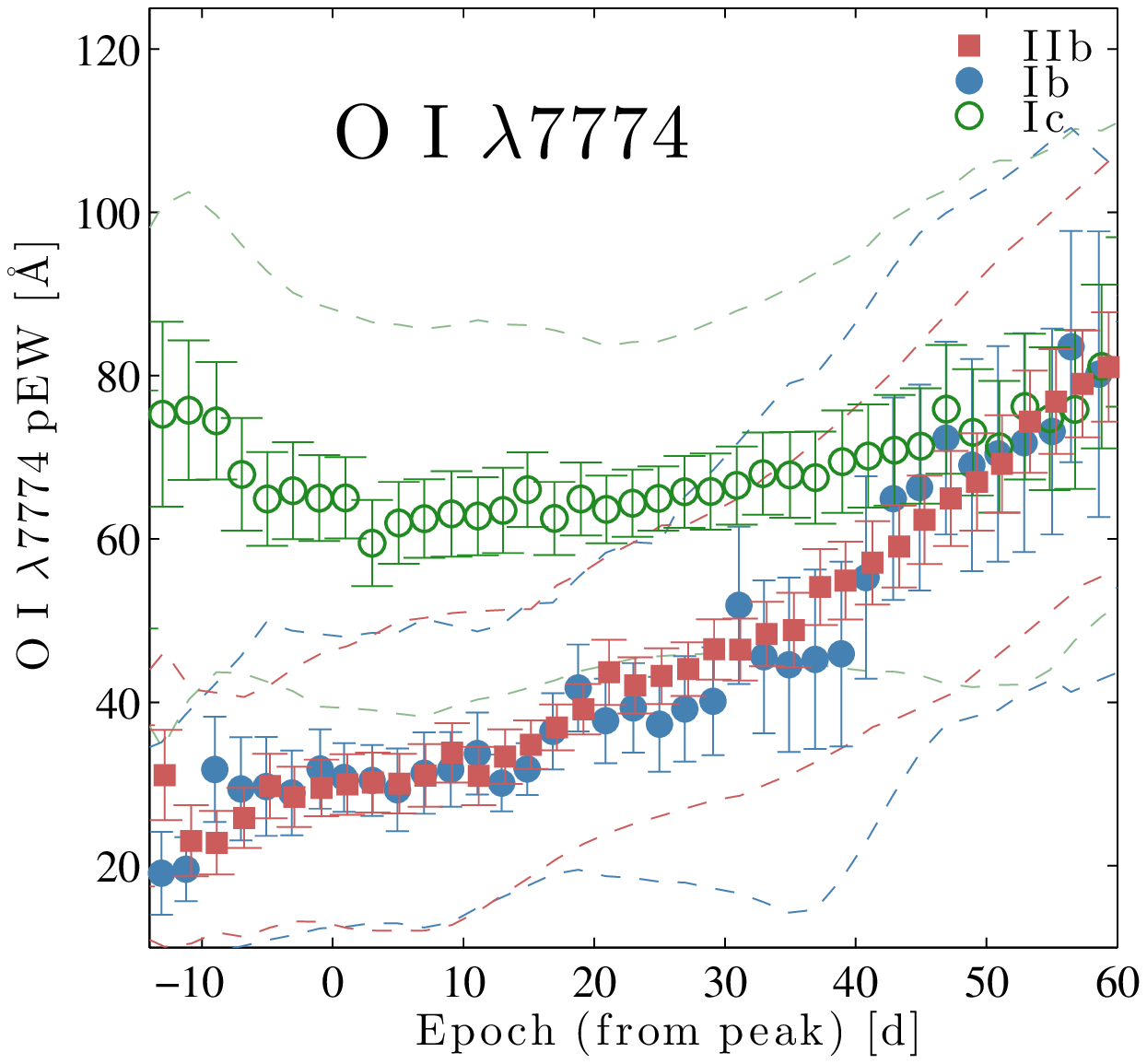}
\\
\includegraphics[width=8cm]{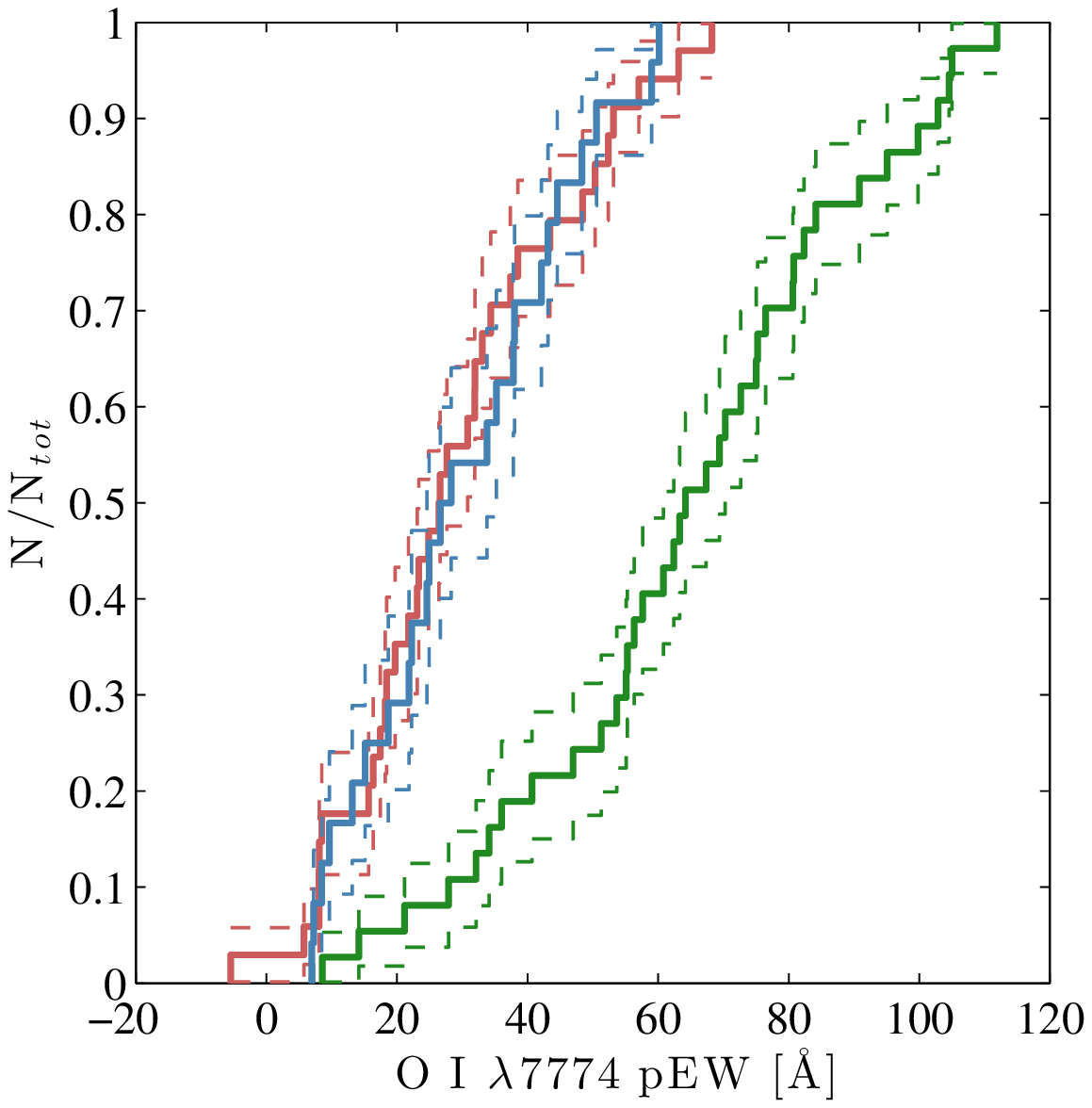}
\caption{Absorption strengths of \ion{O}{i}~$\lambda7774$ (top panels). The top-left panel shows individual measurements, with multiple measurements of the same SN connected by solid lines. The top-right panel shows the averages of the SE SN subtypes, with error bars representing the standard deviation of the mean. Dashed lines outline the standard deviations of the samples for each subtype in matching color. The bottom panel shows CDFs measured between $-10$ and $+10$~d.}
\label{fig:pew_ha_oi}
\end{figure*}

Absorption-strength estimates of the \ion{O}{i}~$\lambda7774$ line are
shown in the top-left panel of Fig.~\ref{fig:pew_ha_oi}. We find that
before $+20$~d, SNe Ic clearly show the highest
pEW values, with no detectable difference between SNe IIb and Ib. The pEW distribution we find for SNe Ic is different from
both our SN IIb and SN Ib distributions, with $p<0.001$ in a
two-component K-S test in both cases, when measured in the interval
$-10$~d to $+10$~d. There is no statistically significant difference
between SNe IIb and Ib for any time interval. The evolution
of the mean value for each class is displayed in the top-right panel of 
Fig.~\ref{fig:pew_ha_oi}, showing that SNe Ic are on average
stronger in this line compared to other SE SNe by almost a factor of
two. The CDFs of the subtypes are shown in the bottom panel of
Fig.~\ref{fig:pew_ha_oi}. Before peak there are very few SNe IIb and Ib with pEW values higher than 60~\AA. Thus, a single pEW
measurement of \ion{O}{i}~$\lambda7774$ on an early-time spectrum can be
used to classify SNe Ic (see also \citealp{2017arXiv170702543S}).

Past $+40$~d, there is no statistically significant difference between the pEW values of any of the SE SN subtypes. However, at these later epochs the measurements become difficult owing to increasingly strong emission lines of [\ion{Ca}{ii}]~$\lambda\lambda7293, 7325$ that affect the continuum typically used to measure the \ion{O}{i}~$\lambda7774$ absorption. The similarity among the subtypes at later epochs could be a result of this contamination. Table~\ref{table:avgs} lists the weighted mean pEW for the \ion{O}{i}~$\lambda7774$ absorption feature for each SE SN subtype at maximum light.

\subsection{Velocities}\label{sec:vel}
Typically, the \ion{Fe}{ii} $\lambda5169$ line is used as a tracer for the expansion velocity of the SN photosphere in SE SNe \citep{Dessart:aa}. The photospheric expansion velocity, measured in this way, can be used for LC modeling
(e.g., \citealp{1982ApJ...253..785A}), and sample studies of SE SN LCs have in general followed this methodology (e.g.,
\citealp{2011ApJ...741...97D}; \citealp{Cano:2013aa};
\citealp{2015A&amp;A...574A..60T}; \citealp{2016MNRAS.457..328L};
\citealp{2016MNRAS.458.2973P}). However, \cite{2015MNRAS.453.2189D} have shown that the notion of a photosphere in SE SNe is ambiguous. Instead, \cite{2016MNRAS.458.1618D} suggest that 
it is more appropriate to estimate the mean expansion rate via the absorption minimum of the \ion{He}{i}~$\lambda5876$ line for SNe IIb and Ib, and the \ion{O}{i}~$\lambda7774$ absorption minimum in SNe Ic. 
Expansion-velocity measurements based on these lines are presented below (average expansion velocities around maximum light can be found in Table~\ref{table:avgs}).

\subsubsection{Helium}\label{sec:helium_vel}
\begin{figure*}[t]
\centering
\vspace{-0.4cm}
\includegraphics[width=7.45cm]{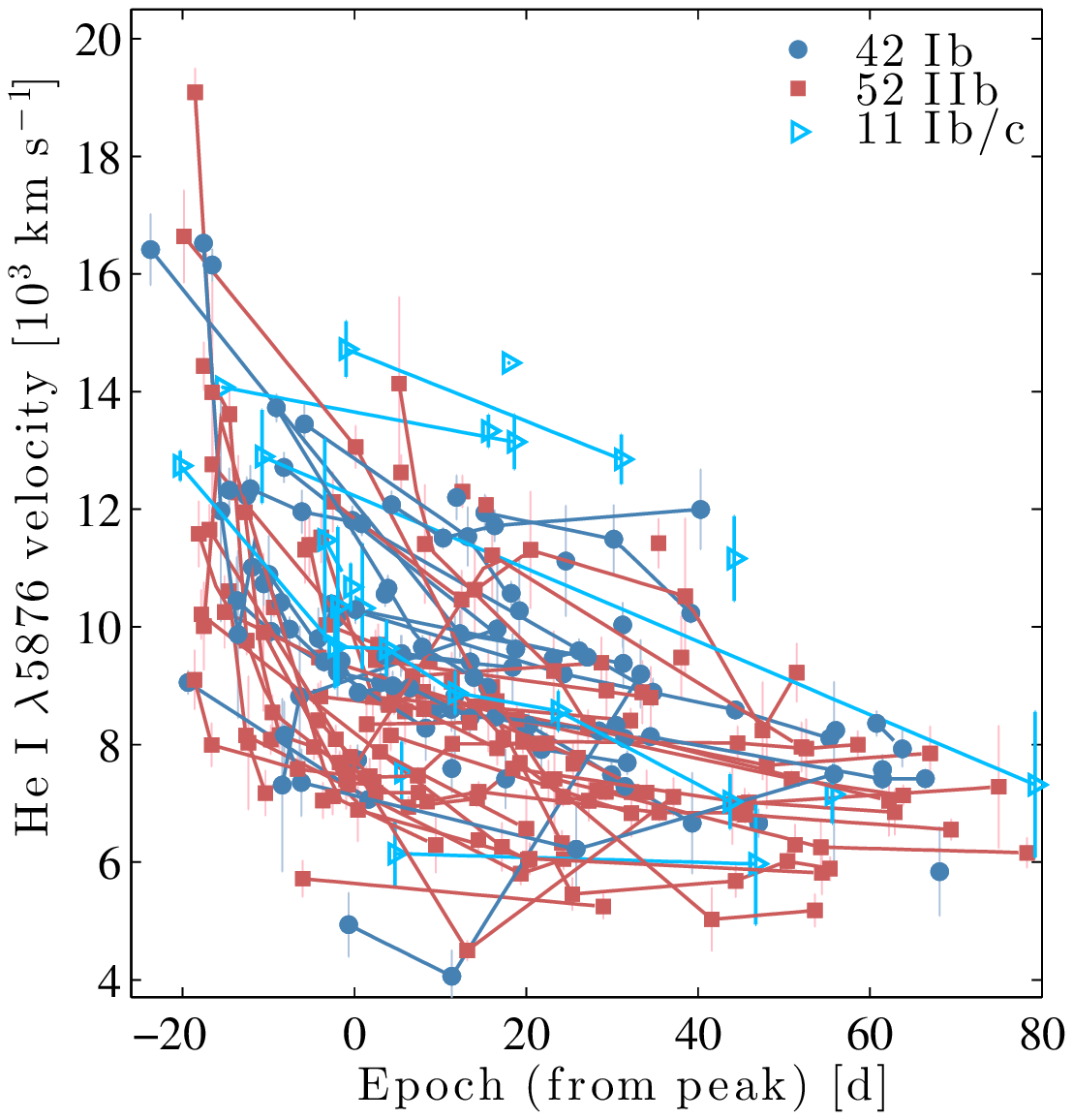}\includegraphics[width=7.55cm,trim=0 0.1cm 0 0]{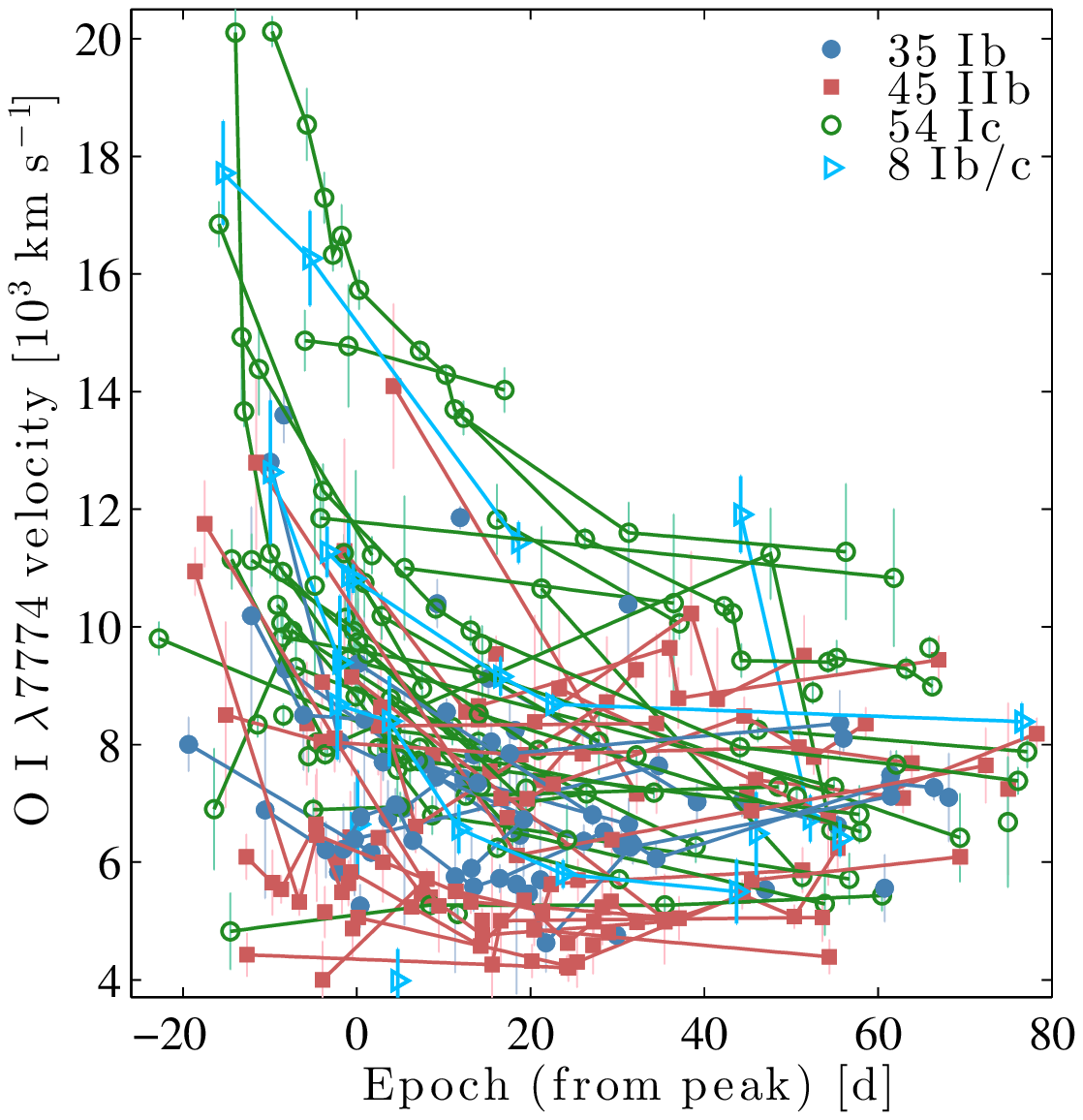}
\\
\vspace{-0.05cm}
\includegraphics[width=7.45cm]{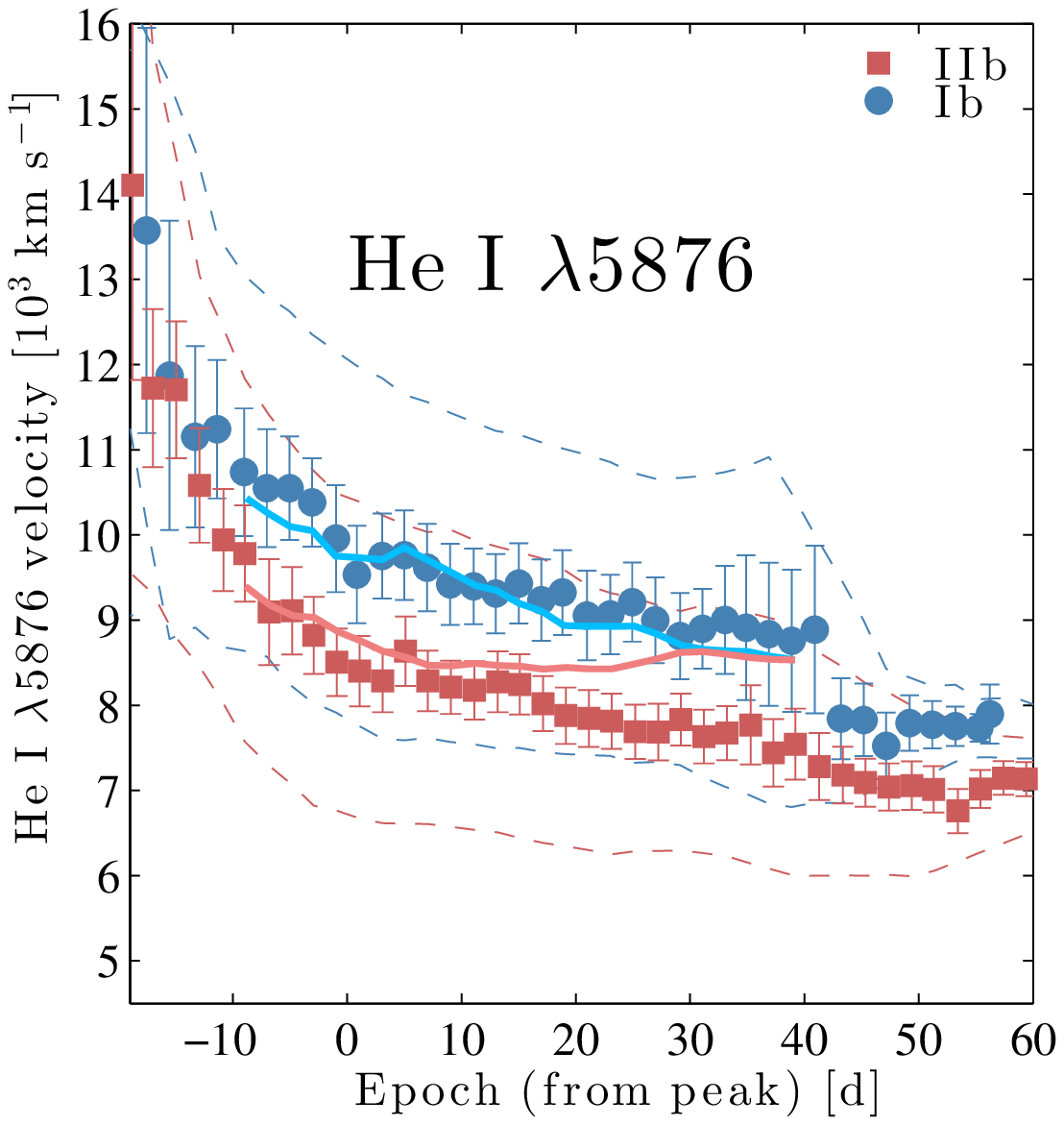}\includegraphics[width=7.55cm,trim=0 0.1cm 0 0]{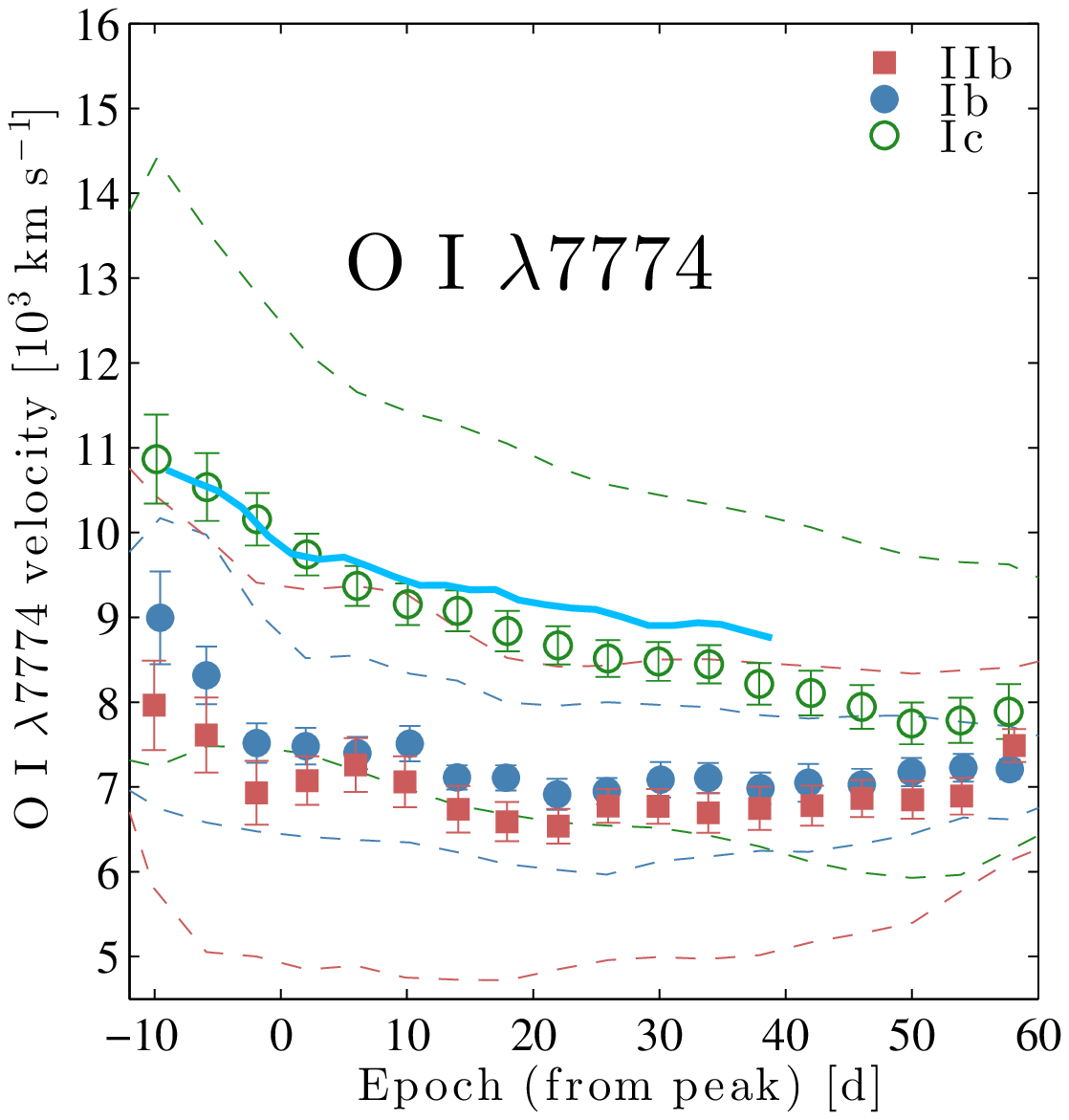}
\\
\vspace{-0.05cm}
\includegraphics[width=7.45cm]{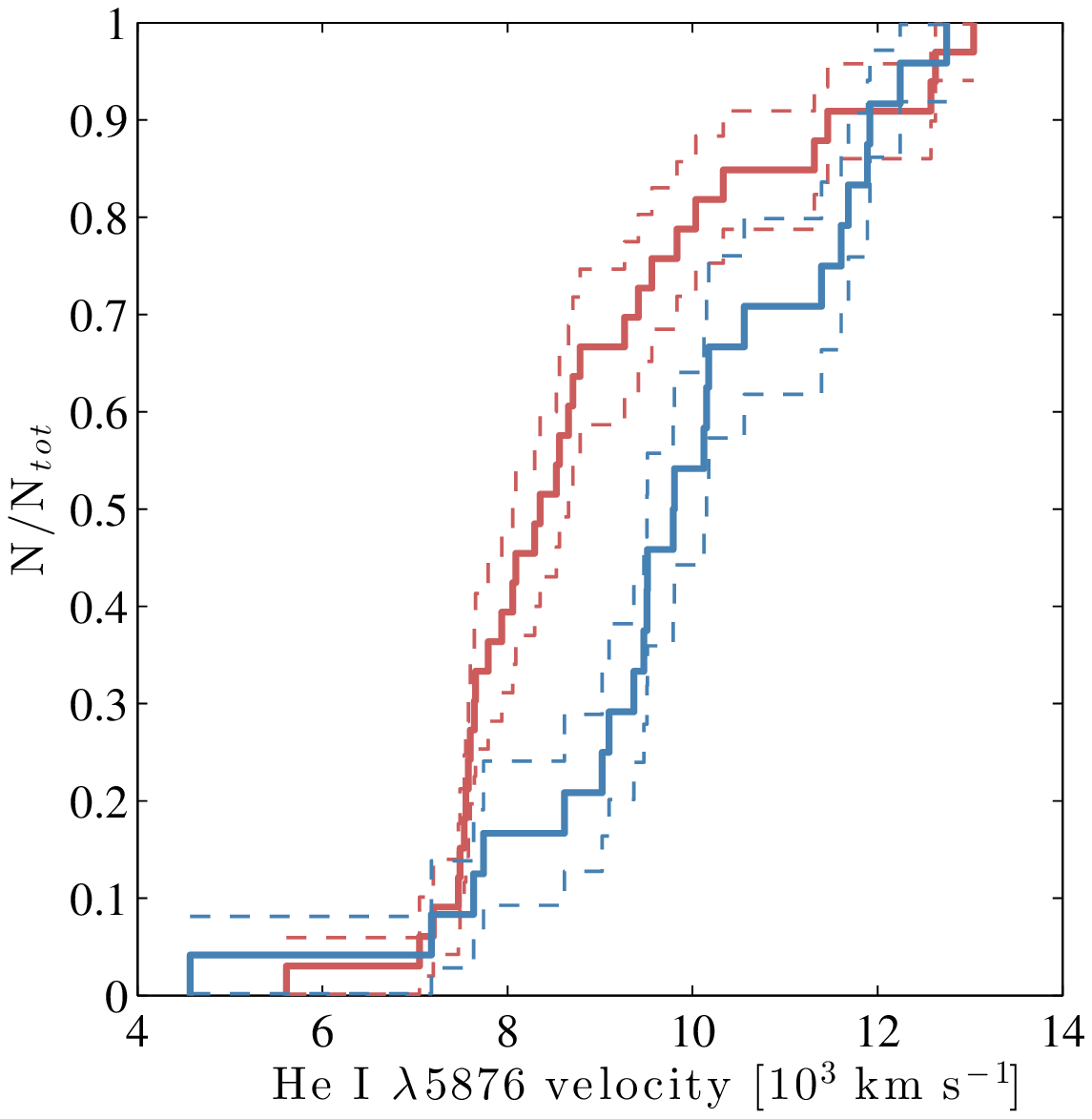}\includegraphics[width=7.45cm,trim=0 0cm 0 0]{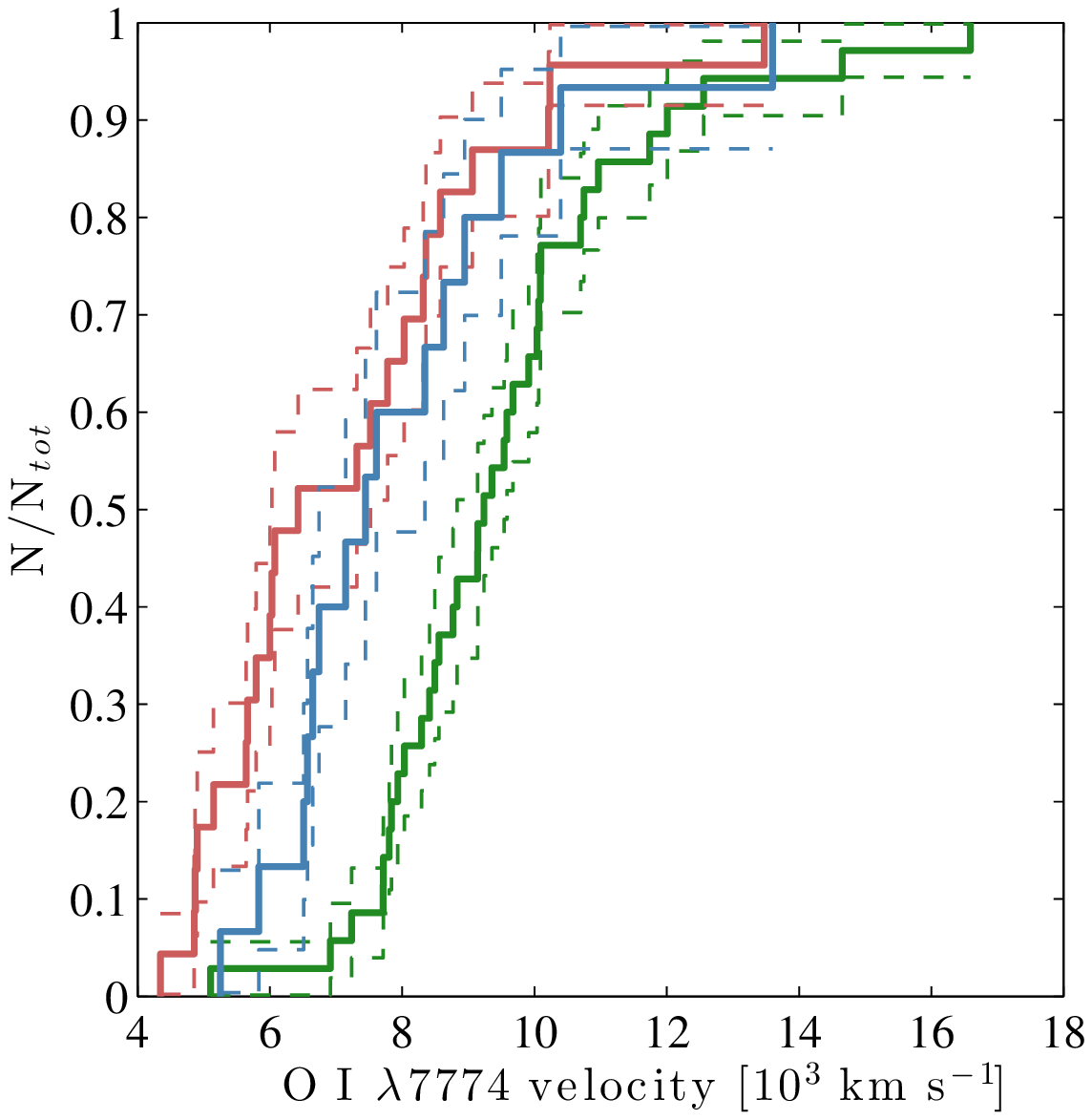}
\vspace{-0.3cm}
\caption{\small{Velocities of \ion{He}{i}~$\lambda5876$ (left panels) and \ion{O}{i}~$\lambda7774$ (right panels). The top panels show individual measurements, with measurements of the same SN connected by solid lines. The middle panels show averages of the SE subtypes, with error bars representing the standard deviation of the mean. Dashed lines outline the standard deviations of the samples for each subtype in matching color. Average velocities derived from \ion{He}{i}~$\lambda7065$ are shown as thick solid lines (Type Ib in blue and Type IIb in red) in the middle-left panel. The average \ion{He}{i}~$\lambda5876$ velocity of Type Ib SNe is shown as a thick solid line in the middle-right panel. The bottom panels show CDFs measured between $-10$ and $+10$~d for each subtype.}}
\label{fig:pew_vels}
\end{figure*}

Expansion-velocity measurements obtained from the minimum of the \ion{He}{i}~$\lambda5876$ absorption line for the SNe IIb and Ib in our sample are shown in the top-left panel of Fig.~\ref{fig:pew_vels}. For SNe Ic, we find that very few objects show velocities that could be consistent with \ion{He}{i}, and there is generally a very large scatter in our measurements (since \ion{He}{} is likely not detected). Thus, we do not include results for SNe Ic in this figure or in Table~\ref{table:avgs}.

SNe Ib tend to be faster, on average, compared to SNe IIb in this line (see the middle-left panel of Fig.~\ref{fig:pew_vels}). Averages obtained from the absorption minimum of the \ion{He}{i}~$\lambda7065$ line (overplotted as thick solid lines in the middle-left panel of Fig.~\ref{fig:pew_vels}) give similar results as for the \ion{He}{i}~$\lambda5876$ line. These findings are consistent with the result by \cite{2016ApJ...827...90L}.

Around maximum light the average \ion{He}{i}~$\lambda5876$ absorption velocity of our SNe Ib is $\sim9500\pm 600$~km~s$^{-1}$. For SNe IIb we measure $8000\pm 500$~km~s$^{-1}$. In a two-component K-S test we find that the SN Ib velocity distribution differs from the SN IIb distribution with $p<0.004$ for the null hypothesis  (calculated between $-10$~d and $+10$~d; see also the CDFs in the bottom-left panel of Fig.~\ref{fig:pew_vels}).

\subsubsection{Oxygen}\label{sec:oxygen_vel}
Velocity estimates based on the absorption minimum of the
\ion{O}{i}~$\lambda7774$ line are shown in the top-right panel of
Fig.~\ref{fig:pew_vels}. We find clear evidence for SNe Ic on
average being faster compared to SNe IIb and Ib, with a few
SNe Ic showing very high velocities at early times, followed by a rapid
decline ($>18,000$~km~s$^{-1}$ before peak; e.g., PTF12gzk;
\citealp{2012ApJ...760L..33B,2013ApJ...778...63H}). A similar result
was found by \cite{2016ApJ...827...90L}. However, their weighted
averages at maximum light were based on only 7 SNe Ic and 4 SNe
Ib. If we consider the time interval of $-2$~d to $2$~d, we have
velocity estimates of 14 SNe IIb, 8 SNe Ib, and 22 SNe Ic. 
In a two-component K-S test for the velocity distributions
between $-10$~d and $+10$~d, we find that SNe Ic differ from both SNe
IIb and Ib with $p<0.01$ for the null hypothesis. Even if we
exclude the 5 fastest SNe Ic in our sample, we still find a
significantly higher average velocity for the remaining SNe Ic
(this can be seen clearly in the CDFs; bottom-right panel of Fig.~\ref{fig:pew_vels}). In the middle-right panel of Fig.~\ref{fig:pew_vels} we also overplot the average \ion{He}{i}~$\lambda5876$ absorption velocity for SNe Ib as a thick solid line. The velocities derived from this line in SNe Ib evolve in a very similar way as the \ion{O}{i}~$\lambda7774$ absorption velocities in SNe Ic.
%insert in figure or in average velocity table how many objects were used for each average for each subtype

Compared to the SNe studied by \cite{2016ApJ...827...90L}, there is somewhat more diversity in our SN Ib sample; we have a few SNe Ib at rather high velocities ($>10,000$~km~s$^{-1}$ around peak brightness). We have also performed velocity measurements of \ion{O}{i}~$\lambda7774$ for our SNe IIb, and find that these tend to be slightly slower compared to our SNe Ib (with a few objects showing velocities $<5000$~km~s$^{-1}$). However, the difference between the SN IIb and SN Ib distributions is not statistically significant in a K-S test ($p<0.19$), when measured between $-10$~d and $+10$~d.

\section{Correlations}\label{sec:correlations}
To further probe the spectral characteristics of the SE SN subtypes, we have searched our sample for correlations among the measured quantities (pEWs and velocities). However, both pEWs and velocities evolve with time. Thus, to minimize the impact of the time dependencies, we restrict ourselves to comparing average values for restricted time intervals. In our search we computed all correlations (Pearson's $r$-values and corresponding $p$-values) for all possible (1~d integer shifted) time bins of $10$~d width starting from $-20$~d to $+50$~d. In all cases where we find statistically significant correlations, we find that at least 10 adjacent time bins give similar correlation coefficients and statistically significant $p$-values. In several cases, the interval $+10$~d to $+20$~d gives particularly clear correlations. 

\begin{figure*}[t]
\centering
\hspace{-0.5cm}
\includegraphics[width=9cm]{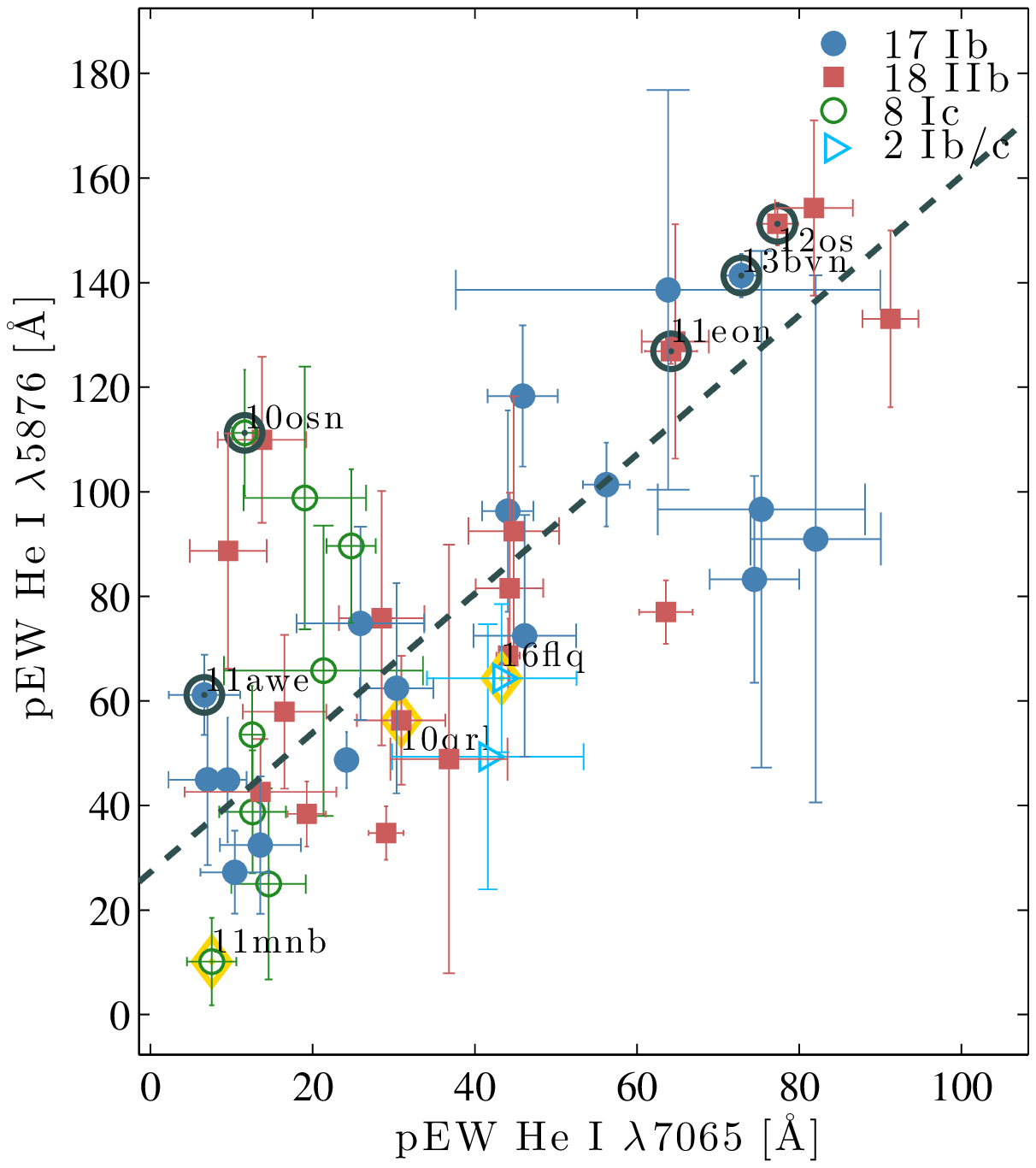}%\includegraphics[width=8cm]{figures2/pEW_types_pew_7774_5876_m10_10d_scatter.eps}
%\\
\includegraphics[width=9.15cm]{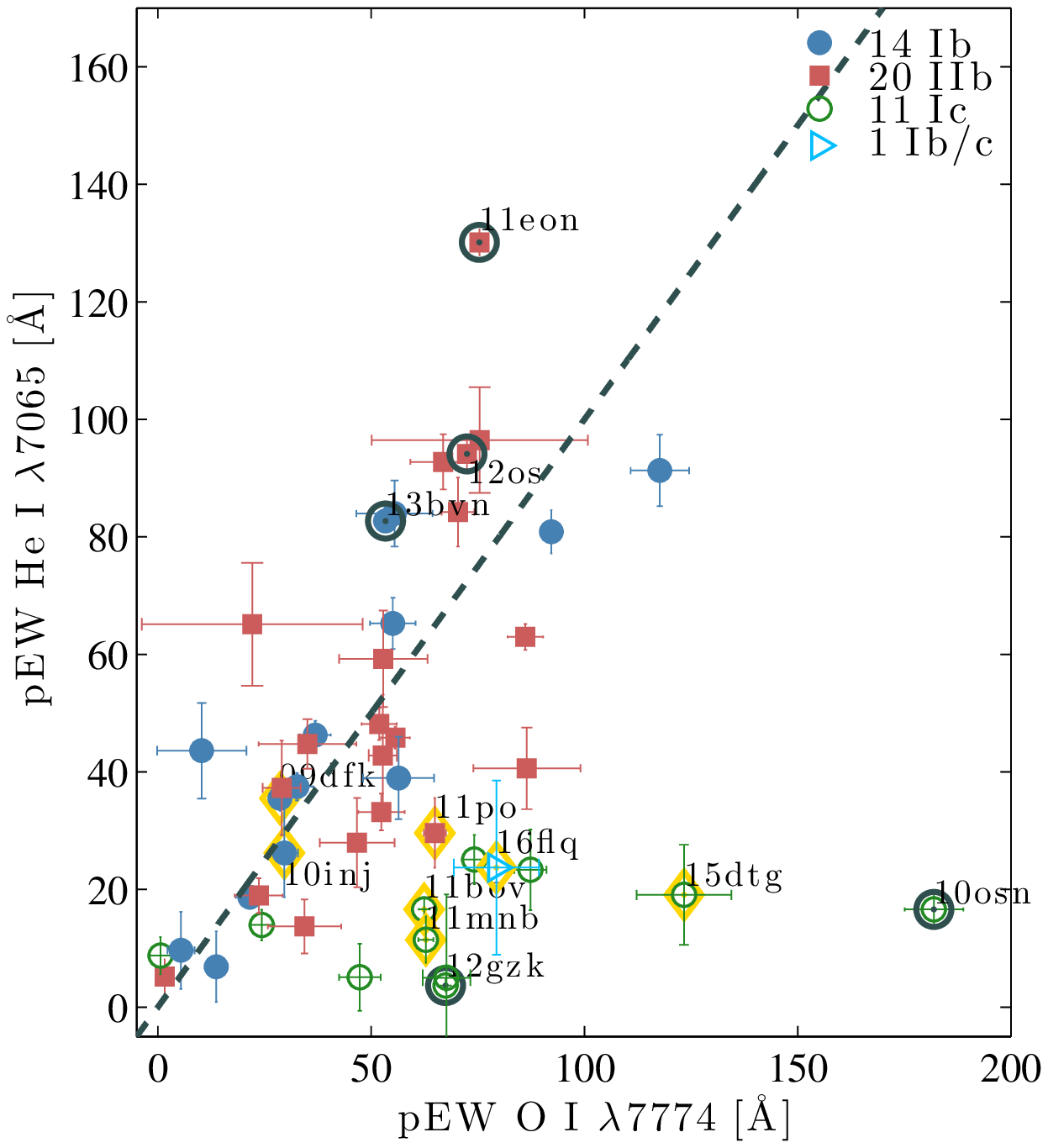}%\includegraphics[width=8cm]{figures2/pEW_types_pew_7774_5876_25_45d_scatter.eps}

\caption{Scatter plots of average pEW values for the SE SNe in our sample. \ion{He}{i}~$\lambda7065$ vs. \ion{He}{i}~$\lambda5876$ calculated between $+10$~d and $+20$~d, along with the best-fit relation $\mathrm{pEW}_{\lambda5876}=30+1.3\times \mathrm{pEW}_{\lambda7065}$ as a dashed line (left panel). \ion{O}{i}~$\lambda7774$ vs. \ion{He}{i}~$\lambda7065$ at $+25$~d to $+45$~d with the best-fit relation $\mathrm{pEW}_{\lambda7065}= 1.0\times \mathrm{pEW}_{\lambda7774}$ (right panel). An illustrative set of SNe have been labeled by their abbreviated (i)PTF names in both panels. Yellow diamonds indicate all objects showing very broad LCs.}
\label{fig:pew_correlations_1}
\end{figure*}

We have also investigated the possible impact of slightly different
temporal evolutions among the SNe. 
Objects with slower evolving LCs could
take more time  
to reach peak absorption strengths, for example, in their \ion{He}{} lines. To
simulate the effect of this, we have stretched the spectral sequence of
SN~2011dh in time, following a Gaussian distribution that corresponds
to the observed LC stretch values of the (i)PTF SE SN sample. 
We find that the potential impact from this effect is not enough to be the main cause of any of our correlations. We find no observational correlations between absorption strength or velocity and LC broadness, except for a few objects with very large stretch values (marked with yellow diamonds in Figs.~\ref{fig:pew_correlations_1} and \ref{fig:pew_correlations_2}). The statistical significances of the correlations shown in Fig.~\ref{fig:pew_correlations_1} and Fig.~\ref{fig:pew_correlations_2} are not affected by including or excluding these objects from the calculations. 

%The potential for host galaxy contamination creating trends in the observed pEW values is difficult to investigate

\subsection{The Helium Shell}\label{sec:corr_helium}
We find that there is a correlation between the pEWs of
\ion{He}{i}~$\lambda7065$ and \ion{He}{i}~$\lambda5876$ (top-left
panel of Fig.~\ref{fig:pew_correlations_1}; measured by computing the
average pEW between $+10$~d and $+20$~d for each SN). There is a large
scatter, but the trend appears to be the same for both SNe IIb and Ib, and we get $r=0.66$ and $p<0.001$, when combining the values
of SNe IIb and Ib. There is no statistically significant
correlation for SNe Ic ($r=0.48$, $p<0.12$). This supports an 
interpretation
where both of these lines are dominated by \ion{He}{} in SNe
IIb and Ib but not in SNe Ic. The latter is also
indicated by the difficulty in obtaining a velocity estimate for
 \ion{He}{i}~$\lambda5876$ for SNe Ic (Sect.~\ref{sec:helium_vel}).
%% OK?
For Type Ic SNe there is a large spread in the pEWs of \ion{He}{i}~$\lambda5876$, but not in \ion{He}{i}~$\lambda7065$. This could indicate that \ion{Na}{i} contamination is quite significant and might be causing most of the scatter (see, e.g., \citealp{2015MNRAS.453.2189D}). A linear fit that excludes SNe Ic gives the relation $\mathrm{pEW}_{\lambda5876}=30(\pm17)+1.3(\pm0.3)\times \mathrm{pEW}_{\lambda7065}$~[\AA], consistent with a 1:1 relation. Furthermore, we find that the trend is similar when calculated for any time interval of similar duration between $-10$~d and $+50$~d. 

In the top-left panel of Fig.~\ref{fig:pew_correlations_2} we show the
average velocity vs. the average pEW of the \ion{He}{i}~$\lambda7065$
absorption line measured between $+10$~d and $+20$~d. For SNe IIb and
Ib there is a decreasing trend, with higher velocities giving
weaker absorptions ($r=-0.63$, $p<0.001$). This is what would be
expected for \ion{He}{} shells of lower mass being accelerated to higher
velocities, assuming similar kinetic energies in the SN explosions.
We find a similar trend for the \ion{He}{i}~$\lambda5876$ line during the
same time interval (not shown). The trend in \ion{He}{i}~$\lambda7065$
is statistically significant when the averages are calculated for all
intervals starting from $-10$~d and ending at $+40$~d. Prior to
$-10$~d, the signatures from \ion{He}{i}~$\lambda7065$ are generally
very weak (see Fig.~\ref{fig:pew_hei}). 
%and \ion{Na}{1} could be
%causing a large scatter in the pEW of the \ion{He}{i}~$\lambda5876$
%line very early on effectively erasing the correlation.

%%?????   Do we know this.   Is the scatter really larger or are you
%%just guessing here?   shorten  cut or rephrase????

\subsection{The Oxygen Shell}
To probe the oxygen-absorbing region of the ejecta, we compare the average pEW calculated between $+25$~d and $+45$~d  for 
\ion{O}{i}~$\lambda7774$ vs. \ion{He}{i}~$\lambda7065$ (right panel of Fig.~\ref{fig:pew_correlations_1}). For both SNe IIb and Ib there is an increasing trend with objects showing stronger \ion{O}{i}~$\lambda7774$ features also having stronger \ion{He}{i}~$\lambda7065$ ($r=0.73$, $p<0.001$ when SNe IIb and Ib are combined). A linear fit, excluding SNe Ic, gives a relation consistent with $\mathrm{pEW}_{\lambda7065}= 1.0\times \mathrm{pEW}_{\lambda7774}$. For the SNe Ic in Fig.~\ref{fig:pew_correlations_1}, we find no statistically significant correlation between the pEWs. The pEW of \ion{He}{i}~$\lambda7065$ is always low, but it can be associated with a wide range of pEWs of \ion{O}{i}~$\lambda7774$. For time intervals earlier than $+20$~d, the trend for SNe IIb and Ib disappears, and around peak brightness the pEWs appear to be randomly scattered. %This could indicate that only a small fraction of the \ion{O}{} present in the ejecta contributes to the observed emission lines at early times in Type IIb and Type Ib SNe. !!!!!!!!!!!!!!!!!!!!! FIX THIS EXPLANATION
 %!!!!!!!!!!!!!!!!!!!!! FIX THIS EXPLANATION

In the top-right panel of Fig.~\ref{fig:pew_correlations_2} we show the average velocity against the average pEW of the \ion{O}{i}~$\lambda7774$ absorption line measured between $+10$~d and $+20$~d. SNe IIb and Ib are scattered with no apparent correlations at lower absorptions and velocities compared to the SNe Ic in our sample (although there are a few outliers among all three subtypes). For SNe Ic there is a decreasing trend, with higher velocities corresponding to weaker absorptions ($r=-0.50$, $p<0.024$). Similarly as what we found for the \ion{He}{} shell, this could indicate that we are observing \ion{O}{} shells of lower mass being accelerated to higher velocities. The lack of such a trend for SNe Ib and IIb would be expected, since the \ion{O}{} absorbing region is located rather deep inside the ejecta behind significant amounts of \ion{He}{} during this phase. 

In the bottom panel of Fig.~\ref{fig:pew_correlations_2} we show the average velocity of the \ion{O}{i}~$\lambda7774$ absorption line vs. the average pEW of the \ion{He}{i}~$\lambda7065$ absorption line, measured between $-10$~d and $+10$~d for SNe IIb, Ib, and Ib/c; for SNe Ic, the average pEW$/2.2$ of \ion{O}{i}~$\lambda7774$ vs. \ion{O}{i}~$\lambda7774$ velocity is shown. SNe IIb and Ib that exhibit fast \ion{O}{} ($>8 000$~km~s$^{-1}$) also show weak \ion{He}{} absorption ($r=-0.65$, $p<0.001$). To show weak \ion{He}{} absorption, we must have faster \ion{He}{} (the top-left panel of Fig.~\ref{fig:pew_correlations_2}). These findings, taken together, suggest that to see strong and fast \ion{O}{} (i.e., the SNe Ic in our sample), the \ion{He}{} envelope must have been physically removed. The trend seen for SNe IIb and Ib is consistent with the trend in pEW vs. velocity of \ion{O}{i}~$\lambda7774$ in SNe Ic, if the pEW is scaled down by a factor of 2.2. This suggests that the stripping does not necessarily stop as the \ion{He}{} envelope is lost; some SNe Ic could be experiencing significant stripping of material from their C-O cores. Higher kinetic energy could also explain the higher velocities in SNe Ic, but it is difficult to explain why this would simultaneously lead to lower pEWs (see Sect.~\ref{sec:discussion} for further discussion).% since there is no statistically significant difference in the observed average kinetic energies among the SE SN subtypes from LC sample studies (which otherwise could explain the velocity differences). %Especially since an increase in kinetic energy by itself (which could explain the velocity differences) does not change the observed subtype of a SE SN (Dessart et al).

\section{Discussion and Future Outlook}
\label{sec:discussion}

The pEWs and velocities we have measured are generally consistent with
those previously reported by \cite{2001AJ....121.1648M},
\cite{2016ApJ...827...90L}, and
\cite{2017MNRAS.469.2672P}\footnote{This study was based largely on
  the same sample as \cite{2016ApJ...827...90L}, and the results are
  consistent between the two papers for the \ion{He}{} line strengths and
  velocities. The \ion{O}{} lines were not analyzed by
  \cite{2017MNRAS.469.2672P}.}. %and our analysis on the whole supports
%a picture where the main difference between the SE subtypes is
%different amounts of physical envelope stripping due to stellar winds
%or binary mass transfer. 
SNe Ic are faster and show stronger
\ion{O}{} pEWs in early-time spectra compared to the other SE SN subtypes, \ion{He}{} pEW is
anticorrelated with \ion{He}{} and \ion{O}{} velocities for SNe IIb and Ib, and \ion{O}{} pEW appears anticorrelated with \ion{O}{} velocity for SNe Ic. LC sample studies show no robust evidence for larger average
ejecta masses or explosion energies in SNe Ic, compared to SNe %%%%%%%%% split up this sentence
IIb and Ib, that could result in faster and stronger oxygen signatures. Therefore, we agree with \cite{2001AJ....121.1648M} and \cite{2016ApJ...827...90L}; in order to have fast and strong \ion{O}{} absorption at early times in most SNe Ic, the simplest explanation would be that we are immediately seeing into their C-O cores; their \ion{H}{} and \ion{He}{} envelopes have been physically removed prior to the explosions due to stellar winds or binary mass transfer. 

However, models by \cite{2012MNRAS.424.2139D} suggest that differences in \Nif\
mixing could give rise to both SNe Ic and Ib, from similar
progenitors with significant \ion{He}{} envelopes ($\sim1.5$~\msun). 
In these models SNe Ic have low mixing, so that the
radioactive \Nif\ is never close enough to 
nonthermally excite the \ion{He}{} in the envelope. Conversely, SNe Ib must 
have highly mixed \Nif, in order to 
display observable \ion{He}{} lines. A consequence of the lower mixing in
these models is narrower and weaker absorption lines, along with lower
velocities measured from absorption minima (especially for the
\ion{O}{i}~$\lambda7774$ absorption; see fig.~13 in
\citealp{2012MNRAS.424.2139D}). Thus, the results by
\cite{2012MNRAS.424.2139D} can be interpreted such that, if different \Nif\ mixing is the main difference between the SE SN
subtypes, SNe Ic should on average show both slower and weaker
\ion{O}{i}~$\lambda7774$ absorption.

Our results are not consistent
with such low mixing models; for the SNe Ic in our sample we
find on average both stronger \ion{O}{i}~$\lambda7774$ absorption
(Fig.~\ref{fig:pew_ha_oi}) and faster velocities
(Fig.~\ref{fig:pew_vels}) at early times. \cite{2016ApJ...827...90L} used a comparable result to argue against low-mixing models for SNe~Ic, with the supporting evidence that LC studies show that SNe Ib and Ic seem to have progenitors of similar mass\footnote{With progenitors of similar mass, the stronger \ion{O}{} signatures in SNe Ic would be explained by a lack of a \ion{He}{} envelope.}. However, if SNe Ic came from higher energy explosions and significantly more massive progenitors compared to SNe~Ib, the result could be higher kinetic energies in the C-O layer, consistent with stronger and faster \ion{O}{} signatures. Furthermore, if the \Nif\ is not significantly mixed into the outer \ion{He}{} envelopes, there will be no \ion{He}{} signatures, and models for the bolometric LC will underestimate the total ejecta masses (and the derived progenitor masses for SNe Ic), since the outer parts of the envelope devoid of \Nif\ would not contribute to the LC (\citealp{2014ApJ...792L..11P}). \cite{2012MNRAS.424.1372A} suggested that~SNe Ic prefer younger stellar populations, which could be consistent with larger progenitor masses ($>30$~\msun). LC studies also suggest that the most massive SE SNe could have higher explosion energies (e.g., \citealp{2016MNRAS.457..328L}). Thus, low-mixing models for SNe Ic cannot easily be ruled out based on just a comparison of average \ion{O}{} strengths and velocities among the subtypes.

%***Chris: Above, you say that younger stellar populations is consistent
% with >30 solar masses. But I thought that at such high masses (perhaps
% above 20 solar masses?), the lifetime before going supernova is 
% essentially independent of mass. Check. Rephrase if necessary?

%There are a few SNe Ic, such as iPTF15dtg in our dataset that do not rise rapidly, and show no He, and have very broad LCs- -- these COULD be a different class of objects, but bolometric LC models do require high mixing...

\begin{figure*}[t]
\centering
\vspace{-0.3cm}
\hspace{-0.5cm}
\includegraphics[width=8.9cm]{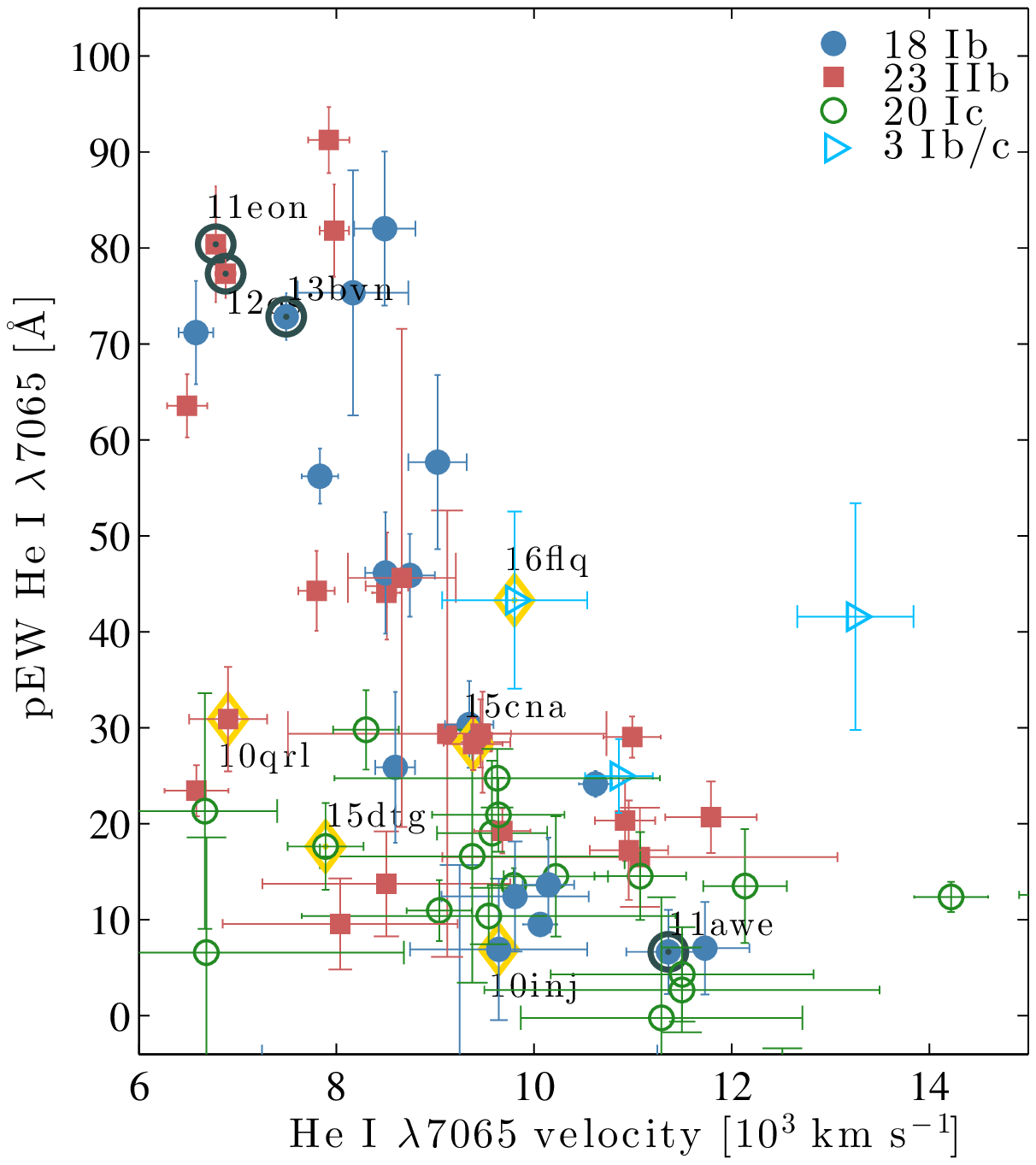}\includegraphics[width=9.05cm]{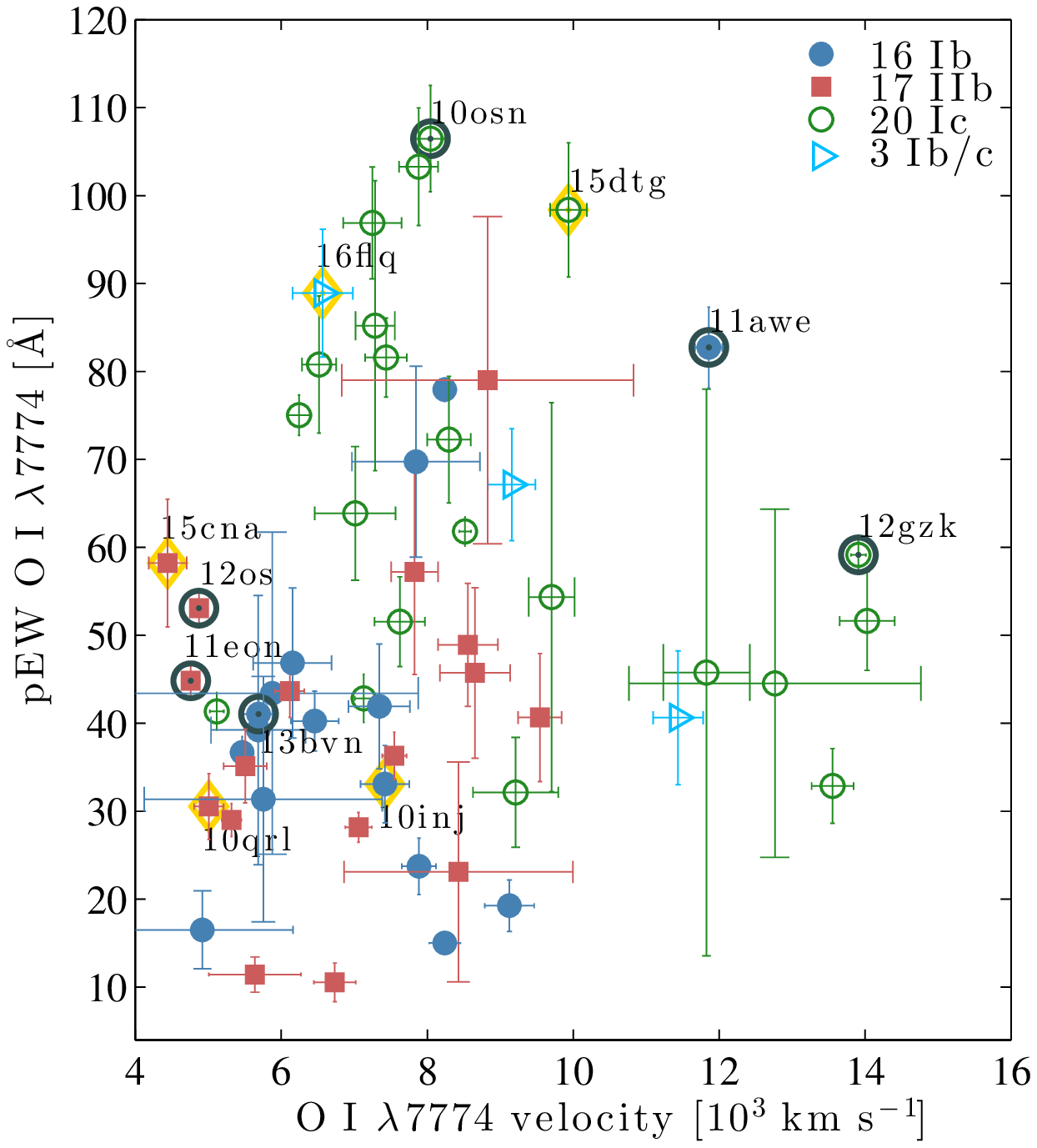}
\\
\hspace{1.5cm}
\includegraphics[width=16cm]{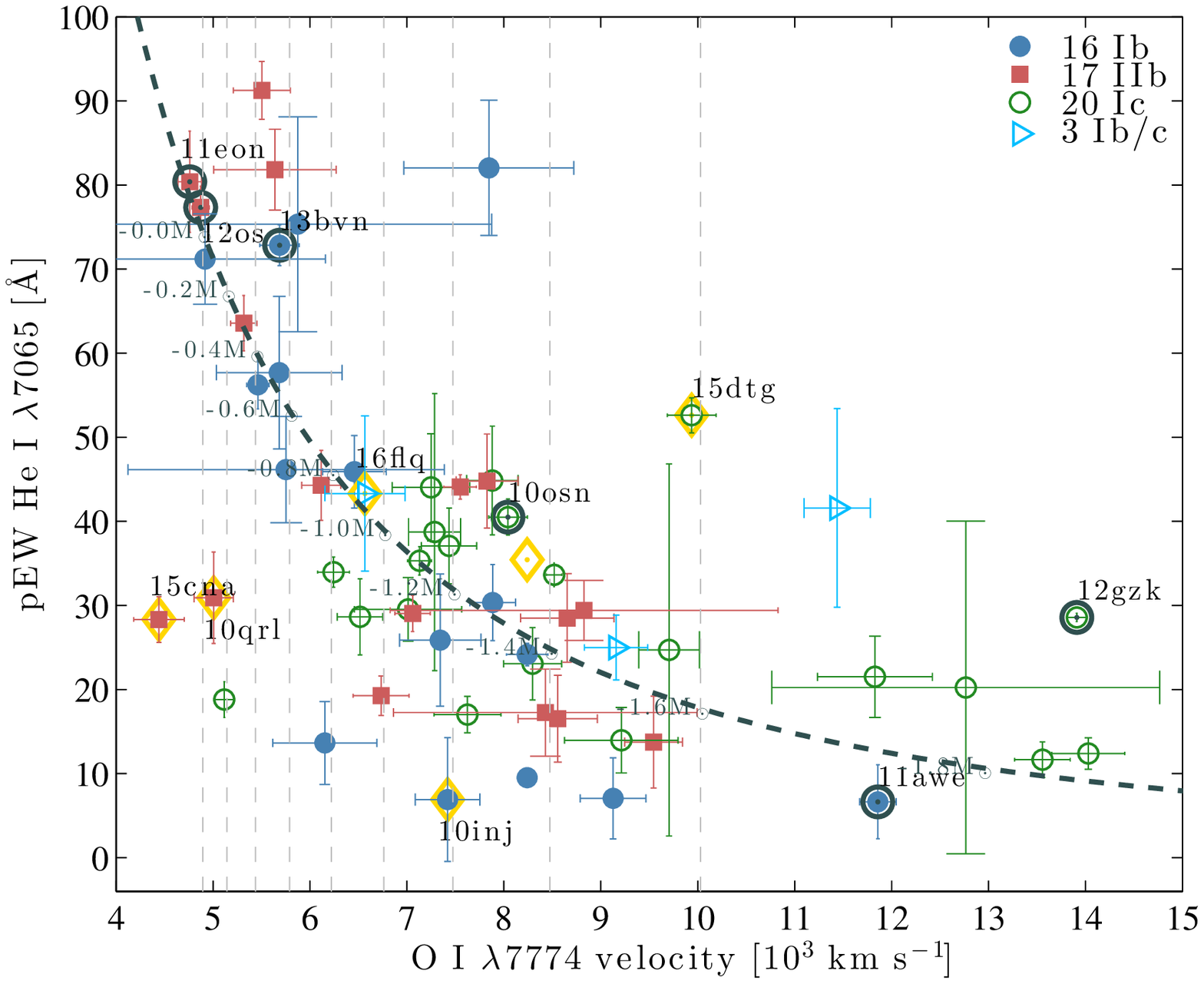}
\vspace{-0.3cm}
\caption{\small{Average pEW vs. velocity of \ion{He}{i}~$\lambda7065$ between $+10$ and $+20$~d (top-left panel). Average pEW vs. velocity of \ion{O}{i}~$\lambda7774$ between $+10$ and $+20$~d (top-right panel). Average pEW of \ion{He}{i}~$\lambda$7065 vs. velocity of \ion{O}{i}~$\lambda$7774 between $+10$~d and $+20$~d for SNe IIb and Ib, and the pEW of \ion{O}{i}~$\lambda7774$ scaled down by a factor of 2.2 vs. \ion{O}{i}~$\lambda7774$ velocity for SNe Ic (bottom panel). An illustrative set of SNe have been labeled by their abbreviated (i)PTF names. Yellow diamonds indicate all objects showing very broad LCs. In the bottom panel the thick dashed black line indicates the best-fit relation pEW$\propto v^{-2}$. The dashed vertical lines indicate the velocity increase expected from the \cite{1982ApJ...253..785A} model by removing $0.2$~\msun\ of material from the ejecta (the \ion{He}{} envelope) for each line from left to right (mass-difference labels follow the thick dashed black line). The model has been scaled, so that the \ion{O}{i}~$\lambda$7774 velocity predicted for a total ejecta mass of $2.1$~\Msun\ is $4900$~km~s$^{-1}$, to match the observed velocities of SN~2011dh (PTF11eon) and PTF12os. Two SNe IIb (iPTF15cna and PTF10qrl) are strong outliers from the fitted trend. These two SNe have very broad LCs and low \ion{O}{i}~$\lambda$7774 velocities (see Sect.~\ref{sec:discussion} for further discussion).}}
\label{fig:pew_correlations_2}
\end{figure*}

To address this issue we looked for correlations among the quantities we have measured (Sect.~\ref{sec:correlations}). In particular, we found that there is an anticorrelation between \ion{He}{} pEW and \ion{O}{} velocity for SNe Ib and IIb, as well as between \ion{O}{} pEW and velocity in SNe Ic (bottom panel, Fig.~\ref{fig:pew_correlations_2}). To explain this behavior with models with gradually increasing kinetic energy (progenitor mass), and \Nif\ mixed in such a way that the \ion{He}{} gradually disappears, would require a significant amount of fine-tuning. The nickel mixing would have to decrease drastically as the kinetic energy increases, which seems counterintuitive since higher velocity ejecta imply higher velocity \Nif. The alternative to this complicated scenario would be that the \ion{He}{} envelope is gradually stripped off --- a much more elegant solution. Thus, we conclude that \ion{He}{} most likely is not hidden in SNe Ic due to low mixing\footnote{See also \cite{2012MNRAS.422...70H}, who have shown that more than $\sim0.1$~\msun\ of He cannot be hidden in highly mixed SE SN models.}. Furthermore, early LC studies of SE SNe indicate that the \Nif\ mixing appears to be high among all SE SN subtypes (\citealp{2015A&amp;A...574A..60T}), and possibly the highest for SNe Ic. Thus, models for normal SNe Ic should have high mixing, similar to the level of mixing needed for good spectral matches to SNe Ib. %\footnote{Note that our results do not exclude the possibility that the kinetic energy could be higher on average in SNe Ic. The bolometric LCs of the SNe must be taken into account to make such statements.}.% (e.g., \citealp{2012MNRAS.424.2139D}). 
%%% ?? I removed Hachinger  unclear if they test the mixing argument??
%%% They only say that not much He can be hidden in their mixed models??

If He cannot be hidden in typical SE SNe due to mixing, the anticorrelation between \ion{He}{} pEW and \ion{O}/\ion{He}{} velocities in Fig.~\ref{fig:pew_correlations_2} suggests that envelopes with a wide range of \ion{He}{} masses or abundances can be present at the time of explosion, since there appears to be a continuum from SNe Ib to Ic in this phase space (top-left panel, Fig.~\ref{fig:pew_correlations_2}). This could be due to progenitors from a range of initial masses undergoing a similar amount of stripping, or a continuum in the amount of stripping experienced by the progenitor stars, or a combination thereof\footnote{Since the transition in \ion{He}{} pEW from SNe Ib to SNe Ic is continuous, there is a range where making the distinction between these subtypes can be ambiguous and dependent on the quality of the spectra (see \citealp{2016ApJ...827...90L,2017PASP..129e4201S}, for further discussion).}. Furthermore, if higher mixing or kinetic energies were the cause of the increasing \ion{O}{i}~$\lambda7774$ velocities from SNe Ib to Ic, there is no clear explanation why the \ion{He}{} lines would disappear as the mixing or kinetic energy increases. For very high velocities ($\gtrsim20,000$~km~s$^{-1}$), spectra do become gradually more featureless (like SNe Ic-BL), and it would be difficult to measure pEWs following our prescription (Sect.~\ref{sec:pewmethod}); but we are here dealing with much lower velocities ($\lesssim10,000$~km~s$^{-1}$). From the models by \cite{2016MNRAS.458.1618D}, it is also clear that varying the kinetic energy, within typical ranges for SE SNe, will not change the observed SN type from a SN Ib to a SN Ic.

%\footnote{Possibly also for \ion{O}{} in Type Ic SNe, however the
%scatter in the data is very high; see the top right panel of
%Fig.~\ref{fig:pew_correlations_2}.}. 
In particular, we find that the 
relation for the pEW of \ion{He}{i}~$\lambda7065$ and the \ion{O}{i}~$\lambda7774$
velocity in SNe IIb and Ib can be fit by pEW$\propto v^{-2}$ (Fig.~\ref{fig:pew_correlations_2}). This suggests that there could be a simple relation between absorption strength and the mass of the \ion{He}{} envelope in SE SNe, since
$M_{\mathrm{ej}} \propto E_\mathrm{k}/v_{\mathrm{ph}}^2$
\citep{1982ApJ...253..785A}. If we assume that $E_{\rm k}$ remains unchanged as the velocity increases, and scale this relation so that the \ion{O}{i}~$\lambda$7774 velocity predicted for a total ejecta mass of $2.1$~\Msun\ is $4900$~km~s$^{-1}$ to match the observed velocities and masses of SN~2011dh (PTF11eon) and PTF12os (\citealp{2016A&amp;A...593A..68F}), we find that to turn these SNe IIb into SNe Ic\footnote{In terms of reaching typical SN Ic \ion{O}{i}~$\lambda$7774 velocities between $+10$~d and $+20$~d, which are 6000 to 10,000~km~s$^{-1}$, for $66.7$\% of the SN Ic population.}, roughly 1.0--1.6~\msun\ of \ion{He}{} would have to be removed (see the dashed vertical lines in the bottom panel of
Fig.~\ref{fig:pew_correlations_2}). This is consistent with the \ion{He}{} envelope mass used in spectral models of SN~2011dh (e.g., \citealp{2015A&amp;A...573A..12J}). Modeling work also clearly shows that a range
of \ion{He}{} envelope masses produces \ion{He}{} signatures of various
strengths (see, e.g., figure 11 in
\citealp{2012MNRAS.422...70H}). However, the exact relationship
between \ion{He}{} absorption pEW and \ion{He}{} envelope mass %for the various lines have 
has not been investigated. We strongly encourage future modeling efforts in this direction. Since the anticorrelation between \ion{He}{} pEW and \ion{O}{} velocity for SNe IIb and Ib is consistent with the behavior of \ion{O}{} pEW and velocity in SNe Ic, our data also support the idea that some SE SNe can experience significant stripping all the way into their C-O cores\footnote{See, e.g., SN~2005ek, possibly the result of mass transfer from a \ion{He}{} star to a compact companion in a very close binary \citep{2013ApJ...774...58D,2013ApJ...778L..23T}.}.%, or iPTF14gqr\footnote{Models for the bolometric LC of iPTF14gqr give very low ejecta mass ($<0.2$~\msun) and low kinetic energy, with an \ion{O}{i}~$\lambda7774$ velocity at maximum light of $>9000$~km~s$^{-1}$.}; De et al. in prep.)}.

\cite{2016MNRAS.458.1618D} have suggested that the best way to constrain the explosion parameters (e.g., total kinetic energy and ejecta mass) for SE SNe is to use the expansion rate of the ejecta ($V_{\rm m}=\sqrt{2E_{\rm k}/M_{\rm ej}}$) derived from the \ion{He}{i}~$\lambda5876$ absorption line for SNe IIb and Ib and from the \ion{O}{i}~$\lambda7774$ line for SNe Ic, instead of $v_{\rm ph}$ in the \cite{1982ApJ...253..785A} model. If we assume that the average kinetic energy among the SN Ib and
Ic subtypes is the same, we can use the observed \ion{He}{i} and \ion{O}{i}
velocities for SNe Ib and Ic at maximum light
(see Table.~\ref{table:avgs} and the middle-right panel of
Fig.~\ref{fig:pew_vels}) to roughly estimate what mass difference
these velocities would imply for these subclasses, which could be interpreted as
the average mass of the \ion{He}{} envelope that is lost in a SN Ic. Using the relation between the ejecta mass ($M_{\mathrm{ej}}$) and the kinetic energy ($E_\mathrm{k}$)  for a constant density sphere expanding homologously (e.g., \citealp{2016MNRAS.457..328L}),
\begin{equation}
\frac{5}{3}\frac{2E_\mathrm{k}}{M_{\mathrm{ej}}}=v_{\mathrm{ph}}^2,
\end{equation}%(Vabs-2640)./0.765
we find with\footnote{$V_{\rm m}$ relations from \cite{2016MNRAS.458.1618D}.} $V_{\mathrm{m-Ib}}=(9500-2640)/0.765\approx9000$~km~s$^{-1}$,
$V_{\mathrm{m-Ic}}=(9800-2990)/0.443\approx15,000$~km~s$^{-1}$, $E_{\rm k}=1\times10^{51}$~erg, and $M_{\rm ej-Ib}=2.1$~\msun, that the $V_{\rm m}$ difference can be reproduced by removing 1.4~\msun\ of
material from the envelope for an average SN Ib (note the similarity to the result based on the \ion{O}{} velocity for SN~2011dh above). However, these calculations hold only if $E_{\rm k}$ really is the same (on average) between the subtypes. This is not necessarily the case; binary models of SE SNe typically require
slightly higher initial masses to produce SNe Ic compared
to SNe Ib (\citealp{2015PASA...32...15Y}). Thus, SNe Ic could have slightly larger C-O cores, and higher kinetic energies, so that the final ejecta masses derived from bolometric LC fitting end up very close to those of SNe Ib. This could affect our estimate of the average \ion{He}{} mass that is lost in a SN Ic, but the anticorrelations we have found between pEW and velocity for both \ion{He}{} and \ion{O}{} cannot easily be explained without removing mass from the envelopes. Spectral modeling efforts in combination with
hydrodynamical modeling are needed in order to make more quantitative statements.

\cite{2014ApJ...792L..11P} suggested that SN~2011dh and
iPTF13bvn could have a significant amount of transparent \ion{He},
based on their low \ion{He}{} absorption velocities (which would result
in underestimated ejecta masses derived from their LCs). We do find
low velocities in these SNe. However, at the same time, they show some
of the strongest \ion{He}{i}~$\lambda\lambda5876, 7065$ absorption
features %among the objects 
in our sample (see, e.g., the bottom panel of Fig.~\ref{fig:pew_correlations_2}). This would not be expected if only a small inner part of the \ion{He}{} envelope is producing the observed \ion{He}{} lines. In the top-left panel of Fig.~\ref{fig:pew_correlations_2} there are only two SNe IIb at velocities lower than $\sim7000$~km~s$^{-1}$ that also show low pEWs. Thus, while this effect could be present in a few objects, it does not appear to be dominating. We also found that SNe Ib and IIb show a trend with high \ion{He}{} pEW values correlated to low velocities (see Fig.~\ref{fig:pew_correlations_2}). This is also inconsistent with partly transparent \ion{He}{} envelopes being typical for these subtypes; we would expect to have more objects at both low pEW and low velocity (only a small part of the \ion{He}{} envelope is emitting, and it should be slow since the emitting region would be located behind an outer transparent envelope of significant mass).

%% ?? check OK with Emir?
There are two SNe IIb (PTF10qrl and iPTF15cna) in our sample that show
weak \ion{He}{i}~$\lambda7065$ absorption, along with very low
\ion{O}{i}~$\lambda7774$ velocities (bottom panel of
Fig.~\ref{fig:pew_correlations_2}). At the same time, their LCs are
very broad (stretched by a factor of $\sim2$, compared to
SN~2011dh; Karamehmetoglu et al., in prep.). This indicates that they likely come from massive
progenitor stars ($\gtrsim30$~\msun). For such progenitors,
even models with high mixing can have difficulties in producing
\ion{He}{} signatures, since the \ion{He}{} envelope is far from the center
of the explosion outside a large C-O core
(\citealp{2015MNRAS.453.2189D}; \citealp{2016MNRAS.458.1618D}). In
these models there is some transparent helium, leading to weak or no
\ion{He}{} signatures, along with low expansion velocities of
\ion{O}{i}~$\lambda7774$ (since the \ion{O}{} is slowed down by a
significant \ion{He}{} envelope), consistent with our observations of
these two SNe. These objects can also be used as an argument against the suggestion by \cite{2013ApJ...773L...7F} that very massive stars ($\gtrsim23$~\msun) would all become SNe Ic with enhanced \ion{O}{} abundances in their outer envelopes owing to strong stellar mixing. The top-right panel of Fig.~\ref{fig:pew_correlations_2} shows that the majority of SE SNe with broad LCs do not exhibit significantly enhanced \ion{O}{} absorption.

% COMMENT ON THE BELOW!!!! important, since we find the continuum!, and add to the main conclusions
 %methods are less prone to errors than the methods used at the time of classification. Independent of data quality or cadence, there is a history of debate in the literature over the exact distinction (if any) between SNe?Ib and SNe?Ic and whether transitional events showing weak helium lines exist (e.g., Filippenko et al. 1990a; Wheeler & Harkness 1990; Wheeler et al. 1994; Clocchiatti et al. 1996; Matheson et al. 2001; Branch et al. 2006).
%The results of recent efforts by Liu & Modjaz (2014), Modjaz et al. (2014), and Liu et al. (2016) argue that the distinction between SNe?Ib and SNe?Ic is useful, and they offer a clearly defined scheme for discriminating between them alongside updated software tools to perform those classifica- tions in a repeatable manner. Modjaz et al. (2014) identify as SNe?Ib all events with detections of both the He I ?6678 and He I ?7065 lines at phases between maximum light and ?50days post-maximum, regardless of line strengths (the stronger He I ?5876 line is also present, but overlaps with Na I). They find that at least one good spectrum observed at these phases is necessary and sufficient to detect the helium lines, which are often absent at pre-maximum and nebular phases even for helium-rich events. Using this classification scheme, they find evidence for a transitional population of ?weak helium? SNe?Ib (Valenti et al. 2011; Modjaz et al. 2014; Liu et al. 2016).

%%%%%%%%%%%%%%%%%

Going forward, we are in the process of modeling the bolometric LCs of all SE SNe found by (i)PTF, using expansion velocities derived from the \ion{He}{i}~$\lambda5876$ line for SNe IIb and Ib and from the \ion{O}{i}~$\lambda7774$ line for SNe Ic. A data-release paper for the full (i)PTF SE SN spectral sample is also in progress, which will supplement the analysis presented here (Fremling et al., in prep.).

%The expansion velocities of our SNe Ib are very similar to our SNe Ic when measured in this way (see Table.~\ref{table:avgs}, and the middle-right panel of Fig.~\ref{fig:pew_vels}). 

%If the ejecta mass is estimated using the velocity of the \ion{He}{i}~$\lambda5876$ for Type IIb and Type Ib SNe and \ion{O}{i}~$\lambda7774$ for Type Ic SNe, we should find no difference in their average ejecta masses assuming that the LCs are of similar widths as in previous SE SN samples.

%Since the LC widths are generally similar among all SE SN subtypes,
%comparable expansion velocities will also likely give similar ejecta masses for the SNe in our sample. 

%The next step is to use the expansion-velocity measurements in this work to model the LCs of the SNe and
%constrain their explosion parameters; such work is in progress. A data-release paper is also in progress, which will supplement the analysis presented here with \ion{H}{} and \ion{Fe}{ii} $\lambda5169$ line diagnostics (Fremling et al., in prep.).

%%%%%%% ADD This could indicate that typically there is a very significant mixing between the \ion{O}{} and \ion{He}{} shells, with the \ion{He}{} shell likely containing a significant fraction of \ion{O}{} for most Type IIb and Type Ib SNe, and lower \ion{He}{} shell mass also results in a lowered amount of \ion{O}. If very different \ion{O}{} fractions would be present in the \ion{He}{} shells of different SNe, one would expect that any pEW of \ion{He}{i} could be associated with both a low and high $pEW$ of \ion{O}{i}

\section{Conclusions}
The main conclusions of this work are as follows.
\begin{itemize}
\item
SNe Ic show higher \ion{O}{i} pEWs and \ion{O}{i} velocities compared to SNe IIb and Ib (higher by $\sim50$~\AA\ and $\sim2000$~km~s$^{-1}$, at maximum light). This is inconsistent with what would be expected if low mixing was responsible for hiding \ion{He}{} signatures in SNe Ic. SNe Ic likely lack \ion{He}{} shells; removing the \ion{He}{} shell will give higher velocities in the \ion{O}{} shell for the same kinetic energy in the explosion.
\item
The \ion{He}{i}~$\lambda\lambda5876, 7065$ velocities at maximum light are higher in SNe Ib compared to SNe IIb. ($\sim1000$~km~s$^{-1}$ higher). The lack of an outer \ion{H}{} shell in SNe Ib allow higher velocities to be reached in the \ion{He}{} shell as the SN shock wave passes through the ejecta.
\item
The pEWs of the \ion{He}{i}~$\lambda\lambda5876, 7065$ absorption lines are similar past maximum light in most SNe Ib and SNe IIb. The \ion{He}{} shells appear to be very similar among SNe IIb and Ib.
\item
There is an anticorrelation between \ion{He}{i}~$\lambda7065$ pEW and \ion{O}{i}~$\lambda7774$ velocity among SNe IIb and Ib. This can be interpreted as a difference in the amount of \ion{He}{} at the time of explosion. The observed difference in the expansion rate of the ejecta around maximum light for SNe Ic and Ib imply an average \ion{He}{} mass difference of $1.4$~\msun, if all other explosion parameters are assumed to be unchanged between SNe Ic and Ib. The difference could either be on account of different starting progenitor masses that undergo a similar amount of stripping, or a continuum in the extent of stripping of the \ion{He}{} envelope from progenitors of similar initial mass, or a combination thereof.
\item
Only two objects in the entire sample (PTF10qrl and iPTF15cna) show both low pEWs of He and slow velocities of either \ion{He}{} or \ion{O}. Both of these objects have very broad LCs, atypical for SE SNe. Therefore, we find that our sample does not show compelling evidence for hidden \ion{He}{} in normal SE SNe. 
%\item
%While there are statistically significant differences in the averages for the SE SN subtypes in most of our measurements, there is always overlap between the standard deviations of the distributions of the three subtypes. There are no gaps in any of our pEW or velocity distributions that could be indicative of clearly different progenitor channels.

\end{itemize}

\section*{Acknowledgements}
\small{
We gratefully  acknowledge support  from  the  Knut and  Alice Wallenberg  Foundation.  The Oskar Klein Centre is funded by the Swedish Research Council. We acknowledge the contributions from the full PTF and iPTF collaborations that made it possible to discover and monitor the SE SNe analyzed in this work. This work was supported by the GROWTH project funded by the National Science Foundation (NSF) under grant AST-1545949. D.A.H. and G.H. are supported by NSF grant AST-1313484. A.V.F. is grateful for financial assistance from NSF grant
AST-1211916, the TABASGO Foundation, the Christopher R. Redlich
Fund, and the Miller Institute for Basic Research in Science (UC Berkeley). His work was conducted in part at the Aspen Center for Physics, 
which is supported by NSF grant PHY-1607611; he thanks the Center for 
its hospitality during the neutron stars workshop in June and July 2017.

Research at Lick Observatory is partially supported by a generous gift 
from Google. Some of the data presented herein were obtained at  the  W. M. Keck
Observatory,  which  is  operated  as  a  scientific  partnership
among the California Institute of Technology, the University of
California, and NASA; the observatory was made possible by the
generous financial support of the W. M. Keck Foundation. 
This work is based in part on observations from the LCO network. 
The William Herschel Telescope is operated on the island of La Palma
by the Isaac Newton Group of Telescopes in the Spanish Observatorio
del Roque de los Muchachos of the Instituto de Astrof\'isica de
Canarias, which is also the site of the Nordic Optical Telescope (NOT) and the Gran Telescopio Canarias (GTC). This work is partly based on observations made with DOLoRes@TNG. Our results made use of the Discovery Channel Telescope (DCT) at Lowell Observatory. Lowell is a private, nonprofit institution dedicated to astrophysical research and public appreciation of astronomy, and it operates the DCT in partnership with Boston University, the University of Maryland, the University of Toledo, Northern Arizona University, and Yale University. The upgrade of the DeVeny optical spectrograph has been funded by a generous grant from John and Ginger Giovale. Initial classification of some SNe was done with the SuperNova Integral Field Spectrograph (SNIFS) on the University of Hawaii 2.2-m telescope as part of the Nearby Supernova Factory II project. We acknowledge the large number of observers and reducers that helped
aquire this spectroscopic database over the years, including 
R. Ellis and M. Sullivan for contributing spectral data obtained under their respective observing programmes. A.V.F. thanks the following members of his group for assistance with
the observations and reductions: J. Choi, R. J. Foley, O. D.  Fox,
M. Kandrashoff, P. J. Kelly, I. Kleiser, J. Kong, A. Miller,
A. Morton, D. Poznanski, and I. Shivvers. We thank the staffs at the observatories where data were obtained.}

%%% add we thank the following people for assistance with observations and reductions :: add form .xls sheet everybody
%\facilities{Keck:I (LRIS), Keck:II (ESI), Hale (DBSP)}

\bibliographystyle{aa}
%\bibliography{ibc_spec}

\clearpage

%
%\begin{figure*}[t]
%\centering
%\includegraphics[width=14cm]{figures2/pEW_types_7065.eps}
%\caption{pEW measurements of the \ion{He}{i} $\lambda\lambda$7065 lines.}
%\label{fig:pew2}
%\end{figure*}
%
%\begin{figure*}[t]
%\centering
%\includegraphics[width=14cm]{figures2/pEW_types_7065_bw.eps}
%\caption{pEW measurements of the \ion{He}{i} $\lambda\lambda$7065 lines.}
%\label{fig:pew3}
%\end{figure*}

%\begin{figure*}[t]
%\centering
%\includegraphics[width=14cm]{figures2/clustering.eps}
%\caption{Clustering measured between $+20$ to $+30$~d of our pEW measurements of the \ion{He}{i} $\lambda\lambda$7065 absorption line.}
%\label{fig:clustering}
%\end{figure*}

%\begin{figure*}[t]
%\centering
%\includegraphics[width=14cm]{figures2/pEW_means_7065.eps}
%\caption{pEW measurements of the \ion{He}{i} $\lambda\lambda$7065 lines.}
%\label{fig:pew}
%\end{figure*}

%\includegraphics[width=8cm]{figures2/pEW_types_Ha.eps}
%\includegraphics[width=8cm]{figures2/pEW_means_Ha.eps}
%\includegraphics[width=8cm]{figures2/pEW_cdf_Ha.eps}Absorption strength measurements of \ion{H}{$\alpha$} (left panels) and \ion{O}{i}~$\lambda$7774~\AA\ (right panels). The top panels show the individual measurements, where measurements of the same objects are connected by dashed lines. The middle panels show the averages of the SE subtypes, with errors representing the uncertainty of the average. The bottom panels show the CDFs measured between $-20$ to $0$~d for \ion{H}{$\alpha$} and $-10$ to $10$~d for \ion{O}{i}~$\lambda$7774~\AA.

%\begin{figure*}[t]
%\centering
%\includegraphics[width=14cm]{figures2/pEW_clustering_Ha.eps}
%\caption{Clustering measured between $-20$ to $0$~d of our pEW measurements of the \ion{H}{$\alpha$} absorption line.}
%\label{fig:clustering_ha}
%\end{figure*}
%[trim=-5cm 0 0 -1cm] `trim=left bottom right top` 

%Scatter plots of \ion{He}{i}~$\lambda$7065~\AA\ vs \ion{H}{$\alpha$} 
% \ion{O}{i}~$\lambda$7774 vs \ion{He}{i}~$\lambda$5876 at $-10$~d and $+10$~d with a dashed vertical line at $pEW_{\lambda7774}=60$~\AA\ (top right panel).

\clearpage

%\tablecomments{Spectra were obtained using the following telescopes and instruments: Keck 1/LRIS \citep{1995PASP..107..375O}, Keck 2/DEIMOS \citep{2003SPIE.4841.1657F}, P200/DBSP \citep{1982PASP...94..586O}, P60/SEDM, Gemini N/GMOS \cite{2004PASP..116..425H}, Gemini S/GMOS-S, WHT/ACAM \citep{2008SPIE.7014E..6XB}, WHT/ISIS, NOT/ALFOSC, ARC 3.5m/DIS, Lick 3m/KAST, FTN/FLOYDS, FTS/FLOYDS, DCT/DeVeny/LMI, GTC/OSIRIS, HET/LRS (ref?), KPNO4m/RCSpec(ref?), Magellan 1/IMACS(ref?), Magellan 2/LDSS3(ref?), SALT/RSS(ref?), TNG/DOLORES(ref?), UH88/SNIFS(ref?), VLT/FORS2(ref?), Wise1m1/FOSC(ref?).}

% insert spectral logs
\begin{onecolumn}
%\begin{centering}
\begin{ThreePartTable}
  \begin{TableNotes}
    \small
      \item\small{Note. --- Spectra were obtained using the following telescopes and instruments: Keck I/LRIS \citep{1995PASP..107..375O}, Keck II/DEIMOS \citep{2003SPIE.4841.1657F}, P200/DBSP \citep{1982PASP...94..586O}, P60/SEDM, Gemini N/GMOS \citep{2004PASP..116..425H}, Gemini S/GMOS-S, WHT/ACAM \citep{2008SPIE.7014E..6XB}, WHT/ISIS, NOT/ALFOSC, ARC 3.5m/DIS, Lick 3m/KAST, FTN/FLOYDS, FTS/FLOYDS, DCT/DeVeny/LMI, GTC/OSIRIS, HET/LRS \citep{1998SPIE.3355..375H}, KPNO 4m/RCSpec, Magellan I/IMACS \citep{2011PASP..123..288D}, Magellan II/LDSS3 \citep{2008SPIE.7014E..0AO}, SALT/RSS, TNG/DOLORES, UH88/SNIFS \citep{2004SPIE.5249..146L}, VLT/FORS2 \citep{1998Msngr..94....1A}, Wise 1m/FOSC.}
      \\
      \item\small{Note 2. --- For objects with a * in the epoch field it was impossible to determine the time of maximum light from their LCs. We have classified these objects based on spectra obtained by the (i)PTF, but they are not included in any further analysis.}
   \end{TableNotes}
\begin{longtable}{lrlrl}
\caption{Spectral log for the (i)PTF SE SN sample.}\label{tab:spec0}\\
%\begin{tabular}{lrlrl}

\toprule
SN & Type & Phase & Redshift & MW$_{E_{B-V}}$ \\ 
{} & {} & [rest-frame days from peak] & $z$ & [mag] \\ 
\midrule
        \endfirsthead
                
        \multicolumn{5}{c}{{\bfseries \tablename\ \thetable{} -- continued from previous
page}}\\
\toprule
SN & Type & Phase & Redshift & MW$_{E_{B-V}}$ \\ 
{} & {} & [rest-frame days from peak] & $z$ & [mag] \\ 
\midrule
        \endhead

\midrule
\multicolumn{5}{r}{{Continued on next page}} \\
\midrule
\endfoot
\bottomrule
\insertTableNotes        \hspace{0.4cm} 
\endlastfoot
   09awk &        Ib &  24 25 & $0.062$ & $0.02$ \\ 
    09dah &       IIb &  -15 -13 19 24 25 44 50 55 & $0.024$ & $0.05$ \\ 
    09dfk &        Ib &  -3 24 44 55 & $0.016$ & $0.05$ \\ 
     09dh &        Ic &  -6 -1 11 17 & $0.070$ & $0.02$ \\ 
    09dha &        Ib &  9 & $0.030$ & $0.02$ \\ 
    09dsj &       IIb &  -1 & $0.135$ & $0.02$ \\ 
    09dxv &       IIb &  -4 2 21 32 & $0.032$ & $0.15$ \\ 
    09dzt &        Ic &  1 & $0.087$ & $0.08$ \\ 
    09fae &       IIb &  5 & $0.067$ & $0.03$ \\ 
    09fsr &        Ib &  18 & $0.008$ & $0.10$ \\ 
    09gxq &       IIb &  16 & $0.038$ & $0.04$ \\ 
    09gyp &       IIb &  -9 & $0.046$ & $0.02$ \\ 
    09hnq &       IIb &  -6 & $0.027$ & $0.09$ \\ 
    09iqd &        Ic &  -5 & $0.034$ & $0.05$ \\ 
    09ism &       IIb &  4 32 & $0.030$ & $0.06$ \\ 
     09ps &        Ic &  4 & $0.106$ & $0.01$ \\ 
     09ut &      Ib/c &  0 & $0.042$ & $0.03$ \\ 
   10abck &        Ib &  -2 4 & $0.014$ & $0.11$ \\ 
   10acbu &        Ib &  46 77 96 & $0.010$ & $0.01$ \\ 
   10acff &        Ib &  -4 & $0.060$ & $0.02$ \\ 
   10acgq &        Ib &  12 41 & $0.105$ & $0.02$ \\ 
    10bhu &        Ic &  6 & $0.036$ & $0.01$ \\ 
    10bip &        Ic &  14 & $0.051$ & $0.02$ \\ 
    10cxx &       IIb &  38 50 & $0.034$ & $0.04$ \\ 
    10eqi &        Ib &  30 & $0.030$ & $0.02$ \\ 
    10fbv &        Ib &  0 & $0.056$ & $0.03$ \\ 
    10feq &        Ib &  -8 19 & $0.028$ & $0.03$ \\ 
    10fia &        Ib &  4 & $0.039$ & $0.05$ \\ 
    10fmr &       IIb &  45 46 63 & $0.020$ & $0.01$ \\ 
    10fmx &        Ic &  21 51 80 & $0.047$ & $0.01$ \\ 
    10fqg &       IIb &  0 & $0.028$ & $0.03$ \\ 
    10gmf &       IIb &  -13 -2 & $0.040$ & $0.02$ \\ 
    10hfe &        Ic &  -7 11 & $0.049$ & $0.02$ \\ 
    10hie &        Ic &  16 37 & $0.067$ & $0.05$ \\ 
     10in &       IIb &  22 51 & $0.070$ & $0.05$ \\ 
    10inj &        Ib &  -24 12 31 68 93 & $0.066$ & $0.01$ \\ 
    10kui &        Ib &  -19 2 & $0.021$ & $0.01$ \\ 
    10lbo &        Ic &  -8 & $0.053$ & $0.01$ \\ 
    10ood &        Ic &  4 5 9 57 & $0.059$ & $0.14$ \\ 
    10osn &        Ic &  -14 -13 -10 4 14 39 & $0.038$ & $0.04$ \\ 
    10pbi &        Ib &  15 30 39 & $0.048$ & $0.07$ \\ 
    10pzp &       IIb &  10 & $0.081$ & $0.02$ \\ 
    10qif &        Ib &  0 35 & $0.064$ & $0.06$ \\ 
    10qqd &        Ic &  0 3 28 & $0.081$ & $0.08$ \\ 
    10qrl &       IIb &  -11 14 & $0.040$ & $0.07$ \\ 
    10svt &        Ic &  75 & $0.031$ & $0.10$ \\ 
    10tqi &        Ic &  -1 0 35 & $0.038$ & $0.03$ \\ 
    10tqv &        Ic &  5 36 65 & $0.080$ & $0.06$ \\ 
    10tud &       IIb &  6 & $0.094$ & $0.02$ \\ 
    10tzh &       IIb &  -18 17 20 & $0.034$ & $0.02$ \\ 
    10vns &       IIb &  34 & $0.040$ & $0.05$ \\ 
    10vnv &        Ib &  47 & $0.015$ & $0.10$ \\ 
    10wal &        Ic &  12 41 & $0.029$ & $0.03$ \\ 
    10wg &      Ib/c &  5 & $0.090$ & $0.01$ \\ 
    10xfl &       IIb &  -1 13 & $0.050$ & $0.09$ \\ 
    10xik &        Ic &  19 63 & $0.071$ & $0.08$ \\ 
    10xjr &        Ib &  31 & $0.030$ & $0.03$ \\ 
    10yow &        Ic &  4 18 19 22 48 54 & $0.025$ & $0.09$ \\ 
    10zcn &        Ic &  14 & $0.020$ & $0.08$ \\ 
    11awe &        Ib &  12 & $0.055$ & $0.02$ \\ 
    11bli &      Ib/c &  5 47 & $0.030$ & $0.01$ \\ 
    11bov &        Ic &  -23 7 34 85 & $0.022$ & $0.03$ \\ 
    11dhf &       IIb &  -4 8 & $0.028$ & $0.02$ \\ 
    11dlg &       IIb &  9 & $0.062$ & $0.03$ \\ 
    11eon &       IIb &  -19 -18 -15 -14 -13 -10 2 7 8 14 35 44 45 69 81 83 91 & $0.002$ & $0.03$ \\ 
    11gcj &      Ib/c &  7 13 & $0.148$ & $0.01$ \\ 
    11hyg &        Ic &  13 49 & $0.030$ & $0.05$ \\ 
    11ilr &        Ib &  20 22 32 & $0.037$ & $0.18$ \\ 
    11ixk &        Ic &  2 & $0.021$ & $0.01$ \\ 
    11izq &        Ib &  5 18 & $0.062$ & $0.01$ \\ 
    11izr &      Ib/c &  -10 44 52 83 & $0.075$ & $0.03$ \\ 
    11jgj &        Ic &  -16 -7 17 & $0.040$ & $0.03$ \\ 
    11kaa &        Ib &  5 13 15 & $0.040$ & $0.02$ \\ 
    11klg &        Ic &  -14 -3 48 55 & $0.027$ & $0.08$ \\ 
    11lmn &      Ib/c &  -5 17 & $0.090$ & $0.04$ \\ 
    11mnb &        Ic &  -15 8 35 60 97 & $0.060$ & $0.02$ \\ 
    11mwk &      Ib/c &  -1 31 & $0.121$ & $0.05$ \\ 
    11pdj &       IIb &  -19 14 36 37 48 67 & $0.024$ & $0.07$ \\ 
    11pnq &      Ib/c &  -15 -5 10 17 19 & $0.074$ & $0.04$ \\ 
     11po &       IIb &  -6 29 & $0.070$ & $0.05$ \\ 
    11prr &       IIb &  -9 -3 12 13 20 38 40 41 47 51 & $0.053$ & $0.12$ \\ 
    11qcj &        Ic &  9 10 12 31 56 66 & $0.028$ & $0.01$ \\ 
    11qiq &        Ib &  -17 -14 -10 8 11 20 39 & $0.032$ & $0.06$ \\ 
    11qju &       IIb &  23 54 78 & $0.028$ & $0.01$ \\ 
    11rka &        Ic &  -1 31 55 & $0.074$ & $0.03$ \\ 
    11rkm &       IIb &  4 8 26 34 88 & $0.065$ & $0.01$ \\ 
    12bwq &        Ib &  -10 -2 3 & $0.040$ & $0.01$ \\ 
    12cde &      Ib/c &  18 46 & $0.013$ & $0.01$ \\ 
    12cjy &        Ic &  0 14 & $0.044$ & $0.01$ \\ 
    12dcp &        Ic &  -9 0 5 18 58 & $0.031$ & $0.02$ \\ 
    12dtf &        Ic &  -9 14 69 & $0.061$ & $0.03$ \\ 
    12eaw &        Ib &  -8 -6 17 & $0.029$ & $0.01$ \\ 
    12eje &       IIb &  9 14 & $0.078$ & $0.01$ \\ 
    12fes &        Ib &  -12 21 & $0.036$ & $0.04$ \\ 
    12fgw &        Ic &  10 & $0.055$ & $0.03$ \\ 
    12fhz &       IIb &  10 43 44 75 & $0.099$ & $0.03$ \\ 
    12fxj &       IIb &  15 & $0.015$ & $0.05$ \\ 
    12gpn &       IIb &  -5 -3 3 7 29 & $0.022$ & $0.03$ \\ 
    12gps &        Ib &  -1 11 33 & $0.016$ & $0.02$ \\ 
    12gty%\tablenotemark{(1)}
     &        Ic &  -11 -6 & $0.176$ & $0.06$ \\ % this needs footnotes, is a Type Ic-BL
    12gvr &      Ib/c &  1 & $0.056$ & $0.03$ \\ 
    12gzk &        Ic &  -16 -15 -12 -11 -10 -6 -4 -3 -2 0 7 10 11 26 42 43 44 54 55 63 66 86 94 95 & $0.014$ & $0.04$ \\ 
    12hni%\tablenotemark{(1)}
     &        Ic &  -5 5 & $0.107$ & $0.05$ \\ 
    12hvv &        Ic &  -12 21 & $0.029$ & $0.07$ \\ 
    12iqw &       IIb &  -17 13 37 & $0.027$ & $0.19$ \\ 
     12jaa &       IIb &  -20 0 24 & $0.024$ & $0.08$ \\ 
    12jxd &        Ic &  14 55 58 86 & $0.025$ & $0.03$ \\ 
    12ktu &        Ic &  -9 19 26 & $0.031$ & $0.06$ \\ 
    12lpo &      Ib/c & * & $0.004$ & $0.03$ \\ 
    12ltw &        Ib &  9 23 26 & $0.060$ & $0.07$ \\ 
    12lvt &        Ib &  61 & $0.012$ & $0.06$ \\ 
    12mfx &        Ib &  0 27 & $0.113$ & $0.06$ \\ 
     12os &       IIb &  -13 -10 -9 -5 -2 -1 2 6 14 19 24 27 28 29 32 93 94 & $0.005$ & $0.04$ \\ 
     13ab &        Ic &  -5 -1 8 & $0.048$ & $0.02$ \\ 
    13aby &       IIb &  -18 -15 -9 -4 19 51 64 83 & $0.018$ & $0.01$ \\ 
    13ajn &       IIb &  -22 -21 -17 2 34 & $0.030$ & $0.04$ \\ 
    13aoo &       IIb &  -4 2 37 & $0.036$ & $0.01$ \\ 
    13aot &        Ic &  66 96 & $0.019$ & $0.01$ \\ 
    13ast &       IIb &  -10 -7 -4 5 17 20 23 45 53 59 81 83 & $0.026$ & $0.01$ \\ 
    13blq &       IIb &  -1 35 & $0.088$ & $0.02$ \\ 
    13bvn &        Ib &  -18 -17 -16 -15 -14 -13 -11 -9 -8 -4 -2 0 4 6 13 16 28 30 31 34 61 66 & $0.004$ & $0.04$ \\ 
    13cab &        Ib &  4 & $0.030$ & $0.06$ \\ 
    13cbf &        Ic &  51 & $0.039$ & $0.06$ \\ 
    13ccj &        Ic &  52 & $0.019$ & $0.05$ \\ 
     13cr &       IIb &  25 & $0.059$ & $0.02$ \\ 
    13cuv &        Ic &  1 & $0.049$ & $0.06$ \\ 
    13dht &        Ic &  -11 & $0.040$ & $0.07$ \\ 
    13djf &        Ic &  -13 -11 2 & $0.021$ & $0.07$ \\ 
    13doq &        Ib &  -4 & $0.072$ & $0.12$ \\ 
    13dug &        Ib &  31 61 & $0.005$ & $0.02$ \\ 
    13ebs &       IIb &  11 17 & $0.027$ & $0.02$ \\ 
    13edf &        Ib &  18 56 & $0.027$ & $0.04$ \\ 
     13nu &       IIb &  -12 9 18 45 72 & $0.026$ & $0.01$ \\ 
      13v &       IIb &  29 & $0.062$ & $0.01$ \\ 
    14aag &       IIb &  1 13 & $0.030$ & $0.03$ \\ 
    14ait &      Ib/c &  56 & $0.039$ & $0.02$ \\ 
    14apl &        Ic &  14 & $0.036$ & $0.03$ \\ 
    14atc &       IIb &  5 8 23 27 62 90 & $0.068$ & $0.01$ \\ 
    14bas &       IIb &  -11 0 20 54 85 & $0.039$ & $0.03$ \\ 
    14bpy &        Ic &  -4 62 64 & $0.045$ & $0.04$ \\ 
    14cyn &        Ib &  -8 -6 26 56 & $0.134$ & $0.03$ \\ 
    14fuz &      Ib/c &  -10 -2 & $0.044$ & $0.02$ \\ 
    14gao &        Ic &  -4 & $0.018$ & $0.07$ \\ 
    14gjv &        Ib &  11 & $0.018$ & $0.02$ \\ 
    14igl &        Ib &  -6 13 19 & $0.039$ & $0.01$ \\ 
    14ikn &       IIb &  -12 -10 16 & $0.020$ & $0.08$ \\ 
    14jhf &      Ib/c &  -11 79 & $0.053$ & $0.02$ \\ 
     14ur &        Ic &  32 & $0.008$ & $0.03$ \\ 
     14va &      Ib/c & * & $0.006$ & $0.02$ \\ 
     14ym &        Ic &  24 & $0.031$ & $0.04$ \\ 
    15acp &        Ic &  -15 -14 2 12 36 & $0.138$ & $0.01$ \\ 
    15acr &       IIb &  -18 -17 -3 9 & $0.061$ & $0.01$ \\ 
    15adv &        Ib &  4 10 16 40 & $0.045$ & $0.02$ \\ 
    15afv &        Ib & * & $0.003$ & $0.08$ \\ 
    15afw &       IIb &  -19 -17 -7 -5 1 7 11 21 22 & $0.008$ & $0.03$ \\ 
    15aiw &        Ib &  -12 -6 1 & $0.067$ & $0.01$ \\ 
    15cam &       IIb &  -1 16 46 52 & $0.028$ & $0.13$ \\ 
    15cna &       IIb &  16 17 23 42 54 83 & $0.060$ & $0.11$ \\ 
    15cpq &      Ib/c &  -3 0 10 17 23 77 & $0.066$ & $0.02$ \\ 
         15dh &        Ic &  -5 & $0.031$ & $0.07$ \\ 
    15dpa &       IIb &  -4 & $0.057$ & $0.02$ \\ 
    15dqb &        Ib &  61 64 & $0.043$ & $0.08$ \\ 
    15dtg &        Ic &  -17 -16 -4 5 9 13 44 62 76 & $0.052$ & $0.05$ \\ 
    15dvg &        Ib &  -9 4 8 14 & $0.047$ & $0.04$ \\ 
    15eoc &       IIb &  3 20 & $0.007$ & $1.16$ \\ 
    15eqv &       IIb & * & $0.005$ & $0.02$ \\ 
    15fhl &        Ic &  16 30 & $0.044$ & $0.03$ \\ 
      15n &        Ib &  10 17 68 & $0.038$ & $0.03$ \\ 
    16ahq &        Ib &  56 & $0.004$ & $0.01$ \\ 
    16bfy &        Ic &  7 & $0.033$ & $0.02$ \\ 
    16flq &      Ib/c &  -20 -2 4 12 24 44 & $0.060$ & $0.02$ \\ 
     16he &       IIb &  -4 14 & $0.031$ & $0.02$ \\ 
    16hgp &        Ic &  -20 -9 & $0.079$ & $0.04$ \\    
\bottomrule
\end{longtable}
%\tablenotetext{(1)}{This SN is a very bright SN Ic, or a faint SLSN (Quimby et al., in prep.). Peak absolute $r$-band is $\sim-20$~mag.}
\end{ThreePartTable}
\end{onecolumn}

\end{document}